\PassOptionsToPackage{unicode}{hyperref}
\PassOptionsToPackage{hyphens}{url}
\documentclass[
  11pt,
]{article}
\usepackage{lmodern}
\usepackage{amssymb,amsmath}
\usepackage{ifxetex,ifluatex}
\ifnum 0\ifxetex 1\fi\ifluatex 1\fi=0 % if pdftex
  \usepackage[T1]{fontenc}
  \usepackage[utf8]{inputenc}
  \usepackage{textcomp} % provide euro and other symbols
\else % if luatex or xetex
  \usepackage{unicode-math}
  \defaultfontfeatures{Scale=MatchLowercase}
  \defaultfontfeatures[\rmfamily]{Ligatures=TeX,Scale=1}
\fi
\IfFileExists{upquote.sty}{\usepackage{upquote}}{}
\IfFileExists{microtype.sty}{% use microtype if available
  \usepackage[]{microtype}
  \UseMicrotypeSet[protrusion]{basicmath} % disable protrusion for tt fonts
}{}
\makeatletter
\@ifundefined{KOMAClassName}{% if non-KOMA class
  \IfFileExists{parskip.sty}{%
    \usepackage{parskip}
  }{% else
    \setlength{\parindent}{0pt}
    \setlength{\parskip}{6pt plus 2pt minus 1pt}}
}{% if KOMA class
  \KOMAoptions{parskip=half}}
\makeatother
\usepackage{xcolor}
\IfFileExists{xurl.sty}{\usepackage{xurl}}{} % add URL line breaks if available
\IfFileExists{bookmark.sty}{\usepackage{bookmark}}{\usepackage{hyperref}}
\hypersetup{
  pdftitle={DynamicLogLog: Faster, Smaller, and More Accurate Cardinality Estimation},
  pdfauthor={Brian Bushnell1*},
  hidelinks,
  pdfcreator={LaTeX via pandoc}}
\usepackage[margin=0.5in]{geometry}
\usepackage{longtable,booktabs}
\usepackage{etoolbox}
\makeatletter
\patchcmd\longtable{\par}{\if@noskipsec\mbox{}\fi\par}{}{}
\makeatother
\IfFileExists{footnotehyper.sty}{\usepackage{footnotehyper}}{\usepackage{footnote}}
\makesavenoteenv{longtable}
\usepackage{graphicx}
\usepackage{caption}
\makeatletter
\def\maxwidth{\ifdim\Gin@nat@width>\linewidth\linewidth\else\Gin@nat@width\fi}
\def\maxheight{\ifdim\Gin@nat@height>\textheight\textheight\else\Gin@nat@height\fi}
\makeatother
\setkeys{Gin}{width=\maxwidth,height=\maxheight,keepaspectratio}
\makeatletter
\def\fps@figure{htbp}
\makeatother
\providecommand{\tightlist}{%
  \setlength{\itemsep}{0pt}\setlength{\parskip}{0pt}}
\newlength{\cslhangindent}
\newenvironment{cslreferences}%
  {}%
  {\par}

\title{DynamicLogLog: Faster, Smaller, and More Accurate Cardinality
Estimation}
\author{Brian Bushnell\textsuperscript{1}*}
\date{2026}

\graphicspath{{figures/}}
\begin{document}
\maketitle

\textsuperscript{1}DOE Joint Genome Institute, Lawrence Berkeley
National Laboratory, Berkeley, CA, USA

*Corresponding author: bbushnell@lbl.gov

ORCiD: Brian Bushnell: https://orcid.org/0000-0002-8140-0131

\hypertarget{abstract}{%
\section{Abstract}\label{abstract}}

Cardinality estimation --- calculating the number of distinct elements
in a stream --- is a longstanding problem with applications across
numerous fields, from networking to bioinformatics to animal population
studies. Widely used approaches include Linear Counting {[}1{]},
accurate at low cardinalities; LogLog {[}2{]}, accurate at high
cardinalities; and HyperLogLog (HLL) {[}3{]}, a fusion of the two, which
is accurate at both low and high cardinalities. HyperLogLog has a
well-known error spike in the mid-region corresponding to the transition
from Linear Counting to LogLog {[}4{]}, though this has been eliminated
in the more recent UltraLogLog (ULL) {[}5{]}. Furthermore, the accuracy
of LogLog improves with the number of buckets (\emph{B}) used (standard
error \(\propto\) \(1/\sqrt{B}\)), and the maximum cardinality it can
represent is limited by the number of bits per bucket. Thus, to increase
the maximum representable cardinality while maintaining accuracy
requires expanding each bucket. Each squaring of the cardinality
(doubling the exponent) requires an additional \emph{B} bits --- the
size of the data structure is \emph{B}\(\times\)log(log(cardinality)).

Here we present DynamicLogLog (DLL), which uses a shared exponent to
allow early exits over \textbf{99.9\%} of the time at high cardinality,
increasing speed while reducing the size complexity to essentially
constant with respect to cardinality, for a chosen precision:
4\emph{B}+log(log(cardinality)). Thus, squaring the maximum
representable cardinality (doubling its exponent) requires only a single
additional bit of global state, regardless of the number of buckets. As
such, while traditional LogLog variants using 6 bits per bucket can
count the stars in the Milky Way, DynamicLogLog can count the particles
in the universe at similar accuracy while using \textbf{33\%} less space
--- and with a flat error curve due to a new blending function.
DynamicLogLog also uses a novel cardinality technique, Dynamic Linear
Counting (DLC), which allows accurate cardinality estimation at any
cardinality without needing a correction factor, as well as a new
Logarithmic Hybrid Blend to eliminate HLL's error hump. DLL's 4-bit
buckets additionally allow more efficient packing in power-of-2 computer
words.

Accuracy was quantified as the width-weighted mean absolute error (each
cardinality interval weighted proportionally), using 2,048 buckets,
averaged over 512,000 simulations out to a true cardinality of
8,388,608, sampled at exponentially-spaced checkpoints. DynamicLogLog's
hybrid estimate demonstrated \textbf{1.83\%} mean and \textbf{1.84\%}
peak absolute error using 1,024 bytes, compared to \textbf{1.83\%}
width-weighted mean and \textbf{3.1\%} peak for HyperLogLog using 1,536
bytes (6 bits \(\times\) 2,048 buckets) --- HLL's higher peak reflects a
transition artifact that DLL eliminates --- while UltraLogLog (ULL)
using 1,024 bytes (8 bits \(\times\) 1,024 buckets) demonstrated
\textbf{1.95\%} mean and \textbf{1.96\%} peak. DLC, which is used to
calculate blend points in DLL's hybrid function, achieved
\textbf{1.87\%} mean and \textbf{1.88\%} peak without correction.
Furthermore, UltraDynamicLogLog (UDLL6), a fusion of DLL and ULL,
achieves ULL-level accuracy at \textbf{75\%} of its memory (1.5 KB vs 2
KB for 2,048 registers). UDLL6's Hybrid LDLC (HLDLC) estimator ---
blending Layered Dynamic Linear Counting with history-corrected hybrid
estimation --- achieves \textbf{1.12\%} log-weighted and \textbf{1.35\%}
width-weighted mean absolute error, outperforming both ULL's FGRA
estimator and reimplementations of Apache DataSketches HLL\_4 {[}6{]}
and HyperLogLogLog {[}7{]} under identical conditions.

\hypertarget{introduction}{%
\section{1. Introduction}\label{introduction}}

Counting the number of distinct elements in a data stream --- the
\emph{cardinality estimation} problem --- arises wherever large datasets
must be summarized in bounded space. Network monitoring systems estimate
the number of distinct IP flows; bioinformatics pipelines count unique
k-mers in sequencing reads to estimate genome size; ecologists can apply
feature analysis to estimate animal populations. In each case, exact
counting requires space proportional to both the cardinality and
per-element information content, which is infeasible for streams of
billions of complex elements.

Probabilistic cardinality estimators trade exactness for dramatically
reduced space. The foundational insight, due to Flajolet and Martin
{[}8{]}, is that the statistical properties of hash values ---
specifically, the lengths of runs of leading zeros --- encode
information about the number of distinct elements seen. This observation
led to a family of increasingly refined algorithms: Probabilistic
Counting {[}8{]}, Linear Counting {[}1{]}, LogLog {[}2{]}, SuperLogLog
{[}2{]}, and HyperLogLog {[}3{]}.

HyperLogLog, the prevailing standard, uses \emph{B} buckets (where
\emph{B} is the number of buckets) of 6 bits each to achieve a standard
error of approximately \(1.04/\sqrt{B}\). At 2,048 buckets (1,536 bytes
packed at 6 bits), this yields roughly 2.3\% standard error (distinct
from mean absolute error) --- sufficient for most applications. However,
HyperLogLog has three well-known limitations:

\begin{enumerate}
\def\labelenumi{\arabic{enumi}.}
\item
  \textbf{The error spike.} HLL uses Linear Counting (LC) for
  cardinalities below \textasciitilde2.5\emph{B} and the harmonic mean
  estimator above \textasciitilde5\emph{B}. In the transition region
  (roughly 2.5\emph{B} to 5\emph{B}), neither estimator is accurate,
  producing a characteristic error spike of approximately 3\% absolute
  error --- roughly 70\% above steady-state levels. The HyperLogLog++
  variant {[}4{]} mitigates this with an empirical bias correction
  table, but does not eliminate it.
\item
  \textbf{Memory scaling.} Each bucket must store the maximum observed
  leading-zero count (NLZ). With 6-bit buckets, the maximum
  representable NLZ is 63, limiting the countable cardinality to
  approximately 2\^{}63 \(\times\) \emph{B}. To square the maximum
  representable cardinality (doubling its exponent), every bucket needs
  an additional bit --- the total data structure size scales as \emph{B}
  \(\times\) log(log(\emph{C})) where \emph{C} is the maximum
  cardinality.
\item
  \textbf{Speed.} Every input element must be hashed, its bucket
  identified, and the stored value compared and potentially updated.
  There is no mechanism to skip elements that cannot possibly affect the
  result.
\end{enumerate}

In this paper we present DynamicLogLog (DLL), a cardinality estimator
that addresses all three limitations simultaneously. DLL uses a
\emph{shared exponent} (called \texttt{minZeros}) across all buckets,
storing only the \emph{relative} leading-zero count per bucket. This
design yields three key benefits:

\begin{itemize}
\item
  \textbf{Smaller.} Only 4 bits per bucket suffice (vs.~6 for HLL), a
  33\% memory reduction. At the same memory budget, DLL can use 50\%
  more buckets, reducing variance by a factor of \(\sqrt{3/2}\)
  \(\approx\) 1.22.
\item
  \textbf{Faster.} The shared exponent enables an \emph{early exit mask}
  (\texttt{eeMask}): a single unsigned comparison against the hash value
  rejects elements whose NLZ is below the current floor. At high
  cardinality, over 99.9\% of elements are rejected before any bucket
  access occurs.
\item
  \textbf{Flatter.} DLL introduces Dynamic Linear Counting (DLC), a
  tier-aware extension of Linear Counting that provides accurate
  estimates across the full cardinality range. Combined with a
  Logarithmic Hybrid Blend using logarithmically scaled mixing weights,
  DLL eliminates the LC-to-LogLog transition spike entirely.
\end{itemize}

DLL's size complexity is 4\emph{B} + log(log(\emph{C})) --- the shared
exponent costs a single global integer, and squaring the maximum
representable cardinality (doubling its exponent) requires only one
additional bit regardless of the number of buckets. In contrast, HLL's
size complexity is 6\emph{B} (or more generally,
\emph{b}\(\times\)\emph{B} where \emph{b} is the bits per bucket, and
\emph{b} must grow with log(log(\emph{C}))).

\begin{figure}
\centering
\includegraphics{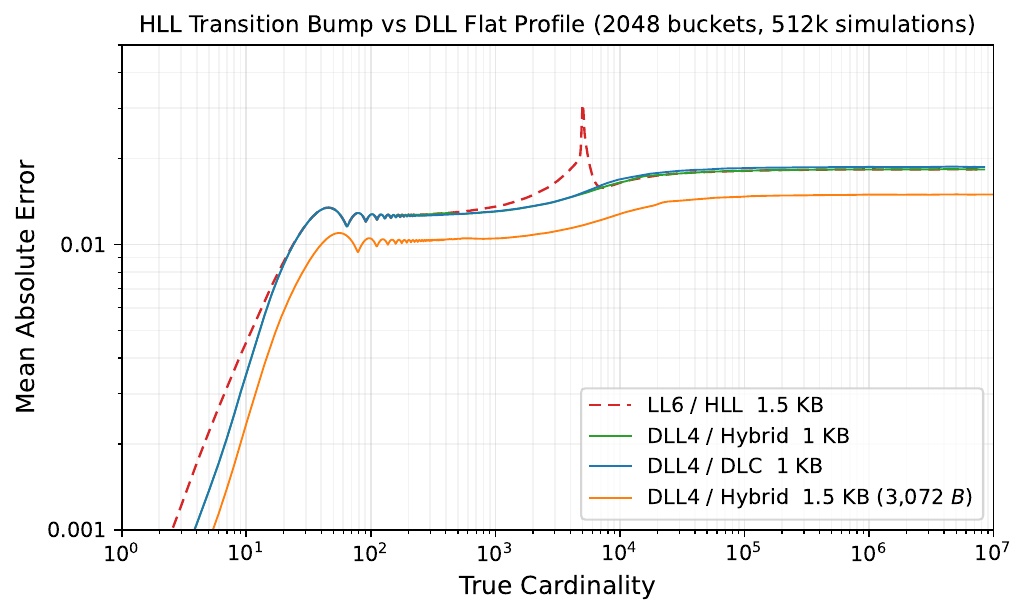}
\caption{\textbf{Figure 1.} Mean absolute error vs true cardinality on a
log-log scale. HLL peaks at 3.1\% error near 2.5\(\times\) \emph{B}
during the LC-to-harmonic-mean transition. DLL Hybrid and DLC remain
below 2\% throughout with no transition artifact. (2048 buckets, 512k
simulations.)}
\end{figure}

\textbf{Related approaches.} Several prior methods reduce per-register
storage using shared offsets. The Apache DataSketches HLL\_4
implementation {[}6{]} stores 4-bit nibble values relative to a global
\texttt{curMin} floor, with an auxiliary \texttt{HashMap} for values
exceeding the 4-bit range --- architecturally similar to DLL4's shared
exponent, but requiring an explicit exception map, with no upper memory
bound. HyperLogLogLog {[}7{]} compresses further to 3-bit registers
relative to a global base, using a sparse exception map for overflows;
periodic rebasing minimizes the exception count. The ``Better with Fewer
Bits'' (tailcut) method {[}9{]} also uses a shared offset with truncated
per-register storage. All three share the core insight of factoring out
a common floor, but differ in how they handle overflow: DLL uses lossy
capping with statistical correction (Section 5.4), while HLL\_4 and HLLL
use exact exception maps. The exception-map approach guarantees
losslessness but adds variable-size auxiliary storage and per-update
overhead, partially negating the memory savings from smaller registers.
ExaLogLog {[}10{]} extends UltraLogLog with variable-width registers and
a Fisher-information-optimal estimator, achieving the best known
memory-variance product among HLL variants at large register counts. The
HyperBitBit family {[}11{]} takes a more radical approach, using only
1--2 bits per element in a flat bit array; like DLL and tailcut, these
estimators are not idempotent (see Section 9).

Ertl {[}12{]} provides a comprehensive treatment of cardinality
estimation methods for HyperLogLog sketches, including a
maximum-likelihood estimator and improved bias corrections that form the
theoretical foundation for FGRA and ExaLogLog.

DLL is implemented in Java as part of the BBTools bioinformatics suite
{[}13{]} and is usable for \emph{k}-mer cardinality counting of sequence
files via the \texttt{loglog.sh} shell script with the flag
\texttt{loglogtype=dll4}. The source code is available in
BBTools/current/cardinality/DynamicLogLog4.java. UDLL6 is in the same
directory as UltraDynamicLogLog6.java.

The remainder of this paper is organized as follows: Section 2 reviews
the mathematical background common to all LogLog-family estimators.
Section 3 describes the DLL architecture, including tier promotion and
the early exit mask. Section 4 introduces the DLC family of estimators
and the Logarithmic Hybrid Blend. Section 5 covers correction factors,
including self-similar correction factor (CF) lookup and the DLL3
overflow correction. Section 6 describes MicroIndex, a 64-bit
cardinality Bloom filter allowing lazy bucket allocation for small sets.
Section 7 describes the simulation and benchmarking methods. Section 8
presents experimental results comparing DLL to HLL on both
high-complexity and low-complexity datasets, evaluates the complementary
nature of DLL and UltraLogLog, introduces history-corrected hybrid
estimation (Hybrid+\emph{n}), presents Layered Dynamic Linear Counting
(LDLC) and its combination with Hybrid+2 (HLDLC), and benchmarks against
external estimator implementations. Section 9 discusses the results and
future directions. Section 10 concludes.

\hypertarget{background}{%
\section{2. Background}\label{background}}

\hypertarget{the-loglog-framework}{%
\subsection{2.1 The LogLog Framework}\label{the-loglog-framework}}

All LogLog-family estimators share a common structure. Given a hash
function h mapping elements to uniformly distributed integers in {[}0,
2\^{}L), the hash value is split into two parts:

\begin{itemize}
\tightlist
\item
  The \textbf{bucket selector}: typically the lowest \emph{k} bits
  determine which of \emph{B} = 2\^{}\emph{k} buckets this element maps
  to (alternatively, a modulo operation allows arbitrary bucket counts).
\item
  The \textbf{rank}: the Number of Leading Zeros (NLZ) in the remaining
  L\(-\)\emph{k} bits. (Flajolet's original definition adds one to avoid
  zero, but DLL uses raw NLZ throughout.)
\end{itemize}

Each bucket stores the \emph{maximum} NLZ observed among all elements
mapping to that bucket. The maximum NLZ encodes information about the
number of distinct elements: if \emph{n} distinct elements are
distributed across \emph{B} buckets, the expected maximum NLZ in each
bucket is approximately log\(_2\)(\emph{n}/\emph{B}).

The key estimators derived from this structure are:

\textbf{Linear Counting (LC).} Counts the fraction of \emph{empty}
buckets. If \emph{V} buckets out of \emph{B} are empty:

\[\text{LC} = B \cdot \ln\!\left(\frac{B}{V}\right)\]

This is accurate when many buckets are empty (cardinality \(\lesssim\)
2.5\emph{B}) but becomes meaningless when all buckets are filled
(\emph{V} = 0).

\textbf{LogLog / HyperLogLog.} Uses the stored maximum NLZ values to
estimate cardinality. The harmonic mean variant (HyperLogLog) computes:

\[\text{HLL} = \alpha_m \cdot B^2 \cdot \left(\sum_{j=1}^{B} 2^{-\text{NLZ}_j}\right)^{-1}\]

where NLZ\_j is the maximum NLZ in bucket \emph{j} (denoted \emph{M\_j}
in Flajolet's original notation) and \(\alpha\)\_m is a bias-correction
constant (\(\approx\) 0.7213/(1 + 1.079/\emph{B}) for large \emph{B}).
This is accurate at high cardinality but unreliable below
\textasciitilde2.5\emph{B}.

\textbf{HLL's transition problem.} HLL uses LC for cardinalities below a
threshold (typically 2.5\emph{B}) and the harmonic mean above it. The
transition between these two estimators produces the characteristic
error spike: neither is accurate in the crossover region, and switching
between them introduces a discontinuity.

\hypertarget{notation}{%
\subsection{2.2 Notation}\label{notation}}

Throughout this paper we use:

\begin{longtable}[]{@{}ll@{}}
\toprule
\begin{minipage}[b]{0.44\columnwidth}\raggedright
Symbol\strut
\end{minipage} & \begin{minipage}[b]{0.50\columnwidth}\raggedright
Meaning\strut
\end{minipage}\tabularnewline
\midrule
\endhead
\begin{minipage}[t]{0.44\columnwidth}\raggedright
\emph{B}\strut
\end{minipage} & \begin{minipage}[t]{0.50\columnwidth}\raggedright
Number of buckets (power of 2 for bitmask selection; arbitrary with
modulo)\strut
\end{minipage}\tabularnewline
\begin{minipage}[t]{0.44\columnwidth}\raggedright
\emph{k}\strut
\end{minipage} & \begin{minipage}[t]{0.50\columnwidth}\raggedright
Bucket selector bits: \emph{B} = 2\^{}\emph{k}\strut
\end{minipage}\tabularnewline
\begin{minipage}[t]{0.44\columnwidth}\raggedright
NLZ\strut
\end{minipage} & \begin{minipage}[t]{0.50\columnwidth}\raggedright
Number of leading zeros in the non-bucket portion of the hash\strut
\end{minipage}\tabularnewline
\begin{minipage}[t]{0.44\columnwidth}\raggedright
absNlz\strut
\end{minipage} & \begin{minipage}[t]{0.50\columnwidth}\raggedright
Absolute NLZ value (the raw count from the hash)\strut
\end{minipage}\tabularnewline
\begin{minipage}[t]{0.44\columnwidth}\raggedright
minZeros\strut
\end{minipage} & \begin{minipage}[t]{0.50\columnwidth}\raggedright
Shared exponent: the current minimum NLZ floor across all buckets
(initialized to $-1$)\strut
\end{minipage}\tabularnewline
\begin{minipage}[t]{0.44\columnwidth}\raggedright
relNlz\strut
\end{minipage} & \begin{minipage}[t]{0.50\columnwidth}\raggedright
Relative NLZ: absNlz \(-\) minZeros\strut
\end{minipage}\tabularnewline
\begin{minipage}[t]{0.44\columnwidth}\raggedright
stored\strut
\end{minipage} & \begin{minipage}[t]{0.50\columnwidth}\raggedright
Encoded bucket value: relNlz (stored = relNlz = absNlz \(-\) minZeros;
empty when absNlz = minZeros + stored = $-1$)\strut
\end{minipage}\tabularnewline
\begin{minipage}[t]{0.44\columnwidth}\raggedright
\emph{V}\strut
\end{minipage} & \begin{minipage}[t]{0.50\columnwidth}\raggedright
Number of empty buckets\strut
\end{minipage}\tabularnewline
\begin{minipage}[t]{0.44\columnwidth}\raggedright
\emph{V\_t}\strut
\end{minipage} & \begin{minipage}[t]{0.50\columnwidth}\raggedright
Number of buckets with absNlz \textless{} \emph{t} (empty at tier
\emph{t})\strut
\end{minipage}\tabularnewline
\begin{minipage}[t]{0.44\columnwidth}\raggedright
\emph{C}\strut
\end{minipage} & \begin{minipage}[t]{0.50\columnwidth}\raggedright
True cardinality\strut
\end{minipage}\tabularnewline
\begin{minipage}[t]{0.44\columnwidth}\raggedright
CF\strut
\end{minipage} & \begin{minipage}[t]{0.50\columnwidth}\raggedright
Correction factor: multiplier applied to a raw estimate\strut
\end{minipage}\tabularnewline
\bottomrule
\end{longtable}

\textbf{Error metrics:}

\begin{longtable}[]{@{}ll@{}}
\toprule
\begin{minipage}[b]{0.40\columnwidth}\raggedright
Metric\strut
\end{minipage} & \begin{minipage}[b]{0.54\columnwidth}\raggedright
Definition\strut
\end{minipage}\tabularnewline
\midrule
\endhead
\begin{minipage}[t]{0.40\columnwidth}\raggedright
Signed error\strut
\end{minipage} & \begin{minipage}[t]{0.54\columnwidth}\raggedright
(estimate \(-\) true) / true. Positive = overcount, negative =
undercount\strut
\end{minipage}\tabularnewline
\begin{minipage}[t]{0.40\columnwidth}\raggedright
Absolute error\strut
\end{minipage} & \begin{minipage}[t]{0.54\columnwidth}\raggedright
\textbar estimate \(-\) true\textbar{} / true. Always non-negative\strut
\end{minipage}\tabularnewline
\begin{minipage}[t]{0.40\columnwidth}\raggedright
Mean absolute error\strut
\end{minipage} & \begin{minipage}[t]{0.54\columnwidth}\raggedright
Average of absolute error across all instances at a given
cardinality\strut
\end{minipage}\tabularnewline
\begin{minipage}[t]{0.40\columnwidth}\raggedright
Standard deviation\strut
\end{minipage} & \begin{minipage}[t]{0.54\columnwidth}\raggedright
Std dev of signed error across instances at a given cardinality;
measures precision\strut
\end{minipage}\tabularnewline
\begin{minipage}[t]{0.40\columnwidth}\raggedright
Peak error\strut
\end{minipage} & \begin{minipage}[t]{0.54\columnwidth}\raggedright
Maximum mean absolute error across all cardinality checkpoints\strut
\end{minipage}\tabularnewline
\begin{minipage}[t]{0.40\columnwidth}\raggedright
Log-weighted avg\strut
\end{minipage} & \begin{minipage}[t]{0.54\columnwidth}\raggedright
Unweighted average of error at exponentially-spaced checkpoints; each
cardinality decade contributes equally. Reflects a log-uniform
cardinality workload\strut
\end{minipage}\tabularnewline
\begin{minipage}[t]{0.40\columnwidth}\raggedright
Card-weighted avg\strut
\end{minipage} & \begin{minipage}[t]{0.54\columnwidth}\raggedright
Each checkpoint weighted by its cardinality; emphasizes high-cardinality
behavior. Reflects a linearly-distributed workload\strut
\end{minipage}\tabularnewline
\bottomrule
\end{longtable}

\textbf{Estimation methods:}

\begin{longtable}[]{@{}ll@{}}
\toprule
\begin{minipage}[b]{0.33\columnwidth}\raggedright
Name\strut
\end{minipage} & \begin{minipage}[b]{0.61\columnwidth}\raggedright
Definition\strut
\end{minipage}\tabularnewline
\midrule
\endhead
\begin{minipage}[t]{0.33\columnwidth}\raggedright
LC\strut
\end{minipage} & \begin{minipage}[t]{0.61\columnwidth}\raggedright
Linear Counting: \emph{B} \(\cdot\) ln(\emph{B}/\emph{V})\strut
\end{minipage}\tabularnewline
\begin{minipage}[t]{0.33\columnwidth}\raggedright
LCmin\strut
\end{minipage} & \begin{minipage}[t]{0.61\columnwidth}\raggedright
Tier-compensated LC: 2\^{}minZeros \(\cdot\) \emph{B} \(\cdot\)
ln(\emph{B}/\emph{V})\strut
\end{minipage}\tabularnewline
\begin{minipage}[t]{0.33\columnwidth}\raggedright
Mean\strut
\end{minipage} & \begin{minipage}[t]{0.61\columnwidth}\raggedright
Occupancy-corrected harmonic mean (Section 5.3)\strut
\end{minipage}\tabularnewline
\begin{minipage}[t]{0.33\columnwidth}\raggedright
HMean\strut
\end{minipage} & \begin{minipage}[t]{0.61\columnwidth}\raggedright
Flajolet's harmonic mean with static \(\alpha\)\_m (Section 5.3)\strut
\end{minipage}\tabularnewline
\begin{minipage}[t]{0.33\columnwidth}\raggedright
GMean\strut
\end{minipage} & \begin{minipage}[t]{0.61\columnwidth}\raggedright
Geometric mean of difference values (Section 5.3)\strut
\end{minipage}\tabularnewline
\begin{minipage}[t]{0.33\columnwidth}\raggedright
HLL\strut
\end{minipage} & \begin{minipage}[t]{0.61\columnwidth}\raggedright
Standard HyperLogLog estimator: \(\alpha\)\_m \(\cdot\) \emph{B}\(^2\)
\(\cdot\) (\(\Sigma\) 2\textsuperscript{(\(-\)NLZ\_j))}($-1$)\strut
\end{minipage}\tabularnewline
\begin{minipage}[t]{0.33\columnwidth}\raggedright
FGRA\strut
\end{minipage} & \begin{minipage}[t]{0.61\columnwidth}\raggedright
Further Generalized Remaining Area estimator (Ertl 2024; Section
8.3)\strut
\end{minipage}\tabularnewline
\begin{minipage}[t]{0.33\columnwidth}\raggedright
DLC(\emph{t})\strut
\end{minipage} & \begin{minipage}[t]{0.61\columnwidth}\raggedright
Dynamic Linear Counting at tier \emph{t}: 2\^{}\emph{t} \(\cdot\)
\emph{B} \(\cdot\) ln(\emph{B}/\emph{V\_t})\strut
\end{minipage}\tabularnewline
\begin{minipage}[t]{0.33\columnwidth}\raggedright
DLC\strut
\end{minipage} & \begin{minipage}[t]{0.61\columnwidth}\raggedright
Exponential log-space blend across all tiers (Section 4.5)\strut
\end{minipage}\tabularnewline
\begin{minipage}[t]{0.33\columnwidth}\raggedright
DLCBest\strut
\end{minipage} & \begin{minipage}[t]{0.61\columnwidth}\raggedright
Best single-tier DLC estimate, selected by optimal occupancy (Section
4.4)\strut
\end{minipage}\tabularnewline
\begin{minipage}[t]{0.33\columnwidth}\raggedright
Hybrid\strut
\end{minipage} & \begin{minipage}[t]{0.61\columnwidth}\raggedright
Logarithmic Hybrid Blend: LCmin + Mean with CF correction (Section
4.6)\strut
\end{minipage}\tabularnewline
\begin{minipage}[t]{0.33\columnwidth}\raggedright
Hybrid+\emph{n}\strut
\end{minipage} & \begin{minipage}[t]{0.61\columnwidth}\raggedright
Hybrid with \emph{n}-bit per-state history correction (e.g., Hybrid+2
uses 2-bit history)\strut
\end{minipage}\tabularnewline
\begin{minipage}[t]{0.33\columnwidth}\raggedright
Mean+\emph{n}\strut
\end{minipage} & \begin{minipage}[t]{0.61\columnwidth}\raggedright
Mean with \emph{n}-bit per-state history correction\strut
\end{minipage}\tabularnewline
\begin{minipage}[t]{0.33\columnwidth}\raggedright
LDLC\strut
\end{minipage} & \begin{minipage}[t]{0.61\columnwidth}\raggedright
Layered Dynamic Linear Counting: per-tier history-corrected LC blend
(Section 8.5)\strut
\end{minipage}\tabularnewline
\begin{minipage}[t]{0.33\columnwidth}\raggedright
HLDLC\strut
\end{minipage} & \begin{minipage}[t]{0.61\columnwidth}\raggedright
Hybrid Layered DLC: 50/50 blend of LDLC and Hybrid+2 (Section 8.5)\strut
\end{minipage}\tabularnewline
\bottomrule
\end{longtable}

\textbf{Estimator types:}

\begin{longtable}[]{@{}lllll@{}}
\toprule
Name & Bits/bucket & Early exit & History bits &
Description\tabularnewline
\midrule
\endhead
LL6 & 6 & No & 0 & Traditional HyperLogLog with unpacked byte
array\tabularnewline
DLL4 & 4 & Yes & 0 & DynamicLogLog, 4-bit\tabularnewline
DLL3 & 3 & Yes & 0 & DynamicLogLog, 3-bit with overflow
correction\tabularnewline
ULL & 8 & No & 2 & UltraLogLog (Ertl 2024)\tabularnewline
UDLL5 & 5 & Yes & 1 & UltraDynamicLogLog, 1-bit history\tabularnewline
UDLL6 & 6 & Yes & 2 & UltraDynamicLogLog, 2-bit history (= DLL + ULL
fusion)\tabularnewline
UDLL7 & 7 & Yes & 3 & UltraDynamicLogLog, 3-bit history\tabularnewline
\bottomrule
\end{longtable}

We differentiate between \emph{Estimator Types} --- the actual data
structure format and method of updating it when a new element arrives
--- and \emph{Estimation Methods}, the algorithm used to derive a
cardinality estimate from the bucket state. These are semi-independent:
many different data structures can yield multiple different estimates.
For example, DynamicLogLog4 (the 4-bit-bucket version) can produce
Linear Counting, Dynamic Linear Counting, HyperLogLog, Mean, and Hybrid
estimates from the same bucket state; similarly, most estimation methods
can be applied to different data structures, sometimes needing minor
modifications (e.g., DLC requires tier-aware bucket counts, which DLL
tracks natively but LL6 can reconstruct from stored NLZ values).

\hypertarget{dynamicloglog-architecture}{%
\section{3. DynamicLogLog
Architecture}\label{dynamicloglog-architecture}}

\hypertarget{shared-exponent-representation}{%
\subsection{3.1 Shared-Exponent
Representation}\label{shared-exponent-representation}}

Like all LogLog-family estimators, DLL requires a hash function that
produces uniformly distributed, unbiased output. Any hash satisfying
this requirement is suitable; our implementation uses Thomas Wang's
64-bit hash {[}14{]}.

The central observation behind DLL is that after enough distinct
elements have been observed, the NLZ values across all \emph{B} buckets
eventually exceed some minimum value. After \emph{n} distinct elements
have been added, the expected maximum NLZ per bucket is approximately
log\(_2\)(\emph{n}/\emph{B}), causing both the average and floor to rise
with increasing cardinality.

Traditional LogLog stores the \emph{absolute} NLZ per bucket, requiring
enough bits to represent the full range (0 to 63 for a 64-bit hash). DLL
instead factors the NLZ into a shared component and a per-bucket
residual:

\[\text{absNlz} = \text{minZeros} + \text{relNlz}\]

The \texttt{minZeros} value is a single integer shared across all
buckets, representing the current ``floor'' --- the minimum NLZ that any
non-empty bucket can have. It is initialized to $-1$, so that empty
buckets (stored = 0) have absNlz = minZeros + stored = $-1$, which is
invalid --- distinguishing them from observed buckets without reserving
a special value. Each bucket stores only the relative NLZ
(\texttt{relNlz}), encoded directly as \texttt{stored\ =\ relNlz}.

With 4 bits per bucket, \texttt{stored} ranges from 0 to 15,
representing relative NLZ values 0 through 15. This 16-tier range is
sufficient: the probability of a valid update causing a bucket to
overflow (relNlz \textgreater{} 15) is approximately 1/65,536 per
update. This is negligible compared to the intrinsic expected error of
the estimator for typical bucket counts (\textasciitilde2,048), though
in practice overflow could be practically eliminated by using 5 bits per
bucket. Thematically, ``approximate algorithms can gain from lossiness
everywhere''.

The memory layout is simple and cache-friendly: 8 buckets pack into a
single 32-bit integer (4 bits \(\times\) 8 = 32), with zero wasted bits.
Compare to 6-bit HLL, where 6 does not divide evenly into 32 or 64,
requiring either wasteful padding or complex cross-word packing.

DynamicLogLog is conceptually similar to a moss or coral. 6-bit
registers can track information at any depth, from the mantle to the
stratosphere. But mosses are short and have no roots; all that matters
is the surface photic zone and immediate surroundings. Rain and dust 1
meter above are irrelevant to the moss until it eventually accumulates
sufficient layers of sediment underneath to reach that height, and rain
that trickles into the dead layers --- below the global floor --- is
unrecoverable.

\hypertarget{tier-promotion}{%
\subsection{3.2 Tier Promotion}\label{tier-promotion}}

As cardinality increases, the NLZ values across all buckets drift
upward. Eventually, every bucket has
\texttt{stored\ \textgreater{}=\ 1}, meaning the shared floor can be
raised. This \emph{tier promotion} is analogous to a floating-point
exponent increment:

\begin{enumerate}
\def\labelenumi{\arabic{enumi}.}
\tightlist
\item
  Increment \texttt{minZeros} by 1.
\item
  Subtract 1 from every bucket's stored value.
\item
  Count the new relNlz=0 buckets to determine \texttt{minZeroCount}.
\item
  If \texttt{minZeroCount} is still 0, repeat (multiple promotions can
  chain).
\end{enumerate}

The complete implementation is shown in Listings 1--2 (after Section
3.3).

\textbf{Merge considerations.} When merging two DLL instances (e.g.,
from parallel threads), the merged result adopts the higher
\texttt{minZeros} and takes the per-bucket max after re-framing.
However, because each instance promoted independently based on its
subset of the data, the merged tier distribution has more overflow than
a single instance would --- see Section 8 (Limitations) for
quantification.

\textbf{Self-similarity.} Each interval between successive tier
promotions --- an \emph{era} --- is statistically self-similar. The
cardinality at which the \emph{t}-th promotion occurs is approximately
\emph{B} \(\times\) 2\^{}\emph{t} \(\times\) ln(\emph{B}), and the
bucket distribution at each promotion boundary is identically shaped
(scaled by a factor of 2). This self-similarity is exploited for
correction factor lookup at arbitrary cardinality (Section 5.2).

\textbf{DLL4} is the primary 4-bit variant of DLL, storing 8 buckets per
32-bit integer. \textbf{DLL3} is a 3-bit variant offering greater memory
savings at slightly reduced accuracy for low-complexity data (Section
5.4). \textbf{LL6} is traditional HyperLogLog with 6 bits per bucket
stored in an unpacked byte array, used throughout this paper as a
baseline for comparison. \textbf{DDL} is a mantissa variant described in
Section 9.

\hypertarget{early-exit-mask}{%
\subsection{3.3 Early Exit Mask}\label{early-exit-mask}}

After \emph{t} tier promotions, all buckets represent NLZ \(\geq\)
\emph{t} and thus any new element whose hash has fewer than \emph{t}
leading zeros cannot possibly update a bucket. Rather than computing the
NLZ and checking, DLL uses a precomputed \emph{early exit mask}
(\texttt{eeMask}) to reject such elements with a single comparison:

\begin{verbatim}
eeMask = 0xFFFFFFFFFFFFFFFFL >>> minZeros   // top minZeros bits cleared
if (key > eeMask) return;                   // unsigned comparison
\end{verbatim}

The fraction of elements that pass the mask is approximately
2\^{}(\(-\)minZeros). Since minZeros \(\approx\)
log\(_2\)(\emph{C}/\emph{B}) where \emph{C} is the current cardinality,
the early exit rate approaches 100\% exponentially --- the higher the
cardinality, the faster DLL becomes per element. This is the origin of
the ``Dynamic'' in DynamicLogLog: the algorithm dynamically adapts its
workload to the data. At low cardinality (minZeros = $-1$), no
elements are rejected and DLL processes every hash like traditional HLL.
As cardinality grows and tiers promote, an exponentially increasing
fraction of elements are rejected before any bucket array memory access.

\textbf{Benchmark.} We measured wall-clock time processing 20 million
150bp genomic sequence reads (2.4 billion 31-mer adds) at a true
cardinality of 1.72 billion (a medium-high complexity stream with 72\%
distinct \emph{k}-mers), using 2048 buckets on a single thread:

\begin{longtable}[]{@{}lll@{}}
\toprule
Configuration & Time & Notes\tabularnewline
\midrule
\endhead
LL6 (HLL baseline) & 23.4 s & Every hash accesses a
bucket\tabularnewline
DLL4 without eeMask & 27.6 s & Tier promotion overhead, no early
exit\tabularnewline
DLL4 with eeMask & 19.6 s & 99.99\% of hashes rejected before bucket
access\tabularnewline
\bottomrule
\end{longtable}

Without eeMask, DLL4 is 18\% \emph{slower} than LL6 due to the overhead
of tier promotion (the \texttt{countAndDecrement} scan at each
promotion). The early exit mask more than compensates: DLL4 with eeMask
is 16\% faster than LL6 overall, despite the tier promotion cost. The
true speedup of eeMask alone is 29\% (27.6s \(\rightarrow\) 19.6s), in a
non-memory-constrained, single-estimator scenario.

At this cardinality (minZeros \(\approx\) 20), only 1 in
\textasciitilde10\(^6\) hashes passes the mask (0.0001\%). Of those that
reach bucket-access logic, only 11.9\% actually update a bucket value.
The vast majority of the 19.6s is spent on \emph{k}-mer generation and
hashing, which occur regardless of the early exit. The actual
bucket-access time is reduced to near zero.

Crucially, the early exit has \emph{zero accuracy impact}: every
rejected element would have produced
\texttt{newStored\ \textless{}=\ oldStored} and been a no-op anyway. The
mask simply avoids the work of confirming this. The eeMask technique is
applicable to any LogLog variant that uses tier promotion, and could be
retroactively added to existing implementations.

\textbf{Amortized cost.} Tier promotions occur only
log\(_2\)(\emph{C}/\emph{B}) times over the lifetime of the estimator
(each time \texttt{minZeroCount} reaches 0). Each promotion runs
\texttt{countAndDecrement} once, looping over all \emph{B} buckets. The
total promotion cost is therefore O(\emph{B} \(\cdot\)
log(\emph{C}/\emph{B})), which is negligible relative to the O(\emph{C})
cost of processing all elements --- particularly since eeMask reduces
the per-element cost to near zero at high cardinality. (Note: the
\emph{storage} cost of \texttt{minZeros} is only log(log(\emph{C}))
bits, since that many bits suffice to represent the promotion count; the
distinction between the number of promotions and the bits needed to
count them is the difference between time and space complexity.)

\begin{figure}
\centering
\includegraphics{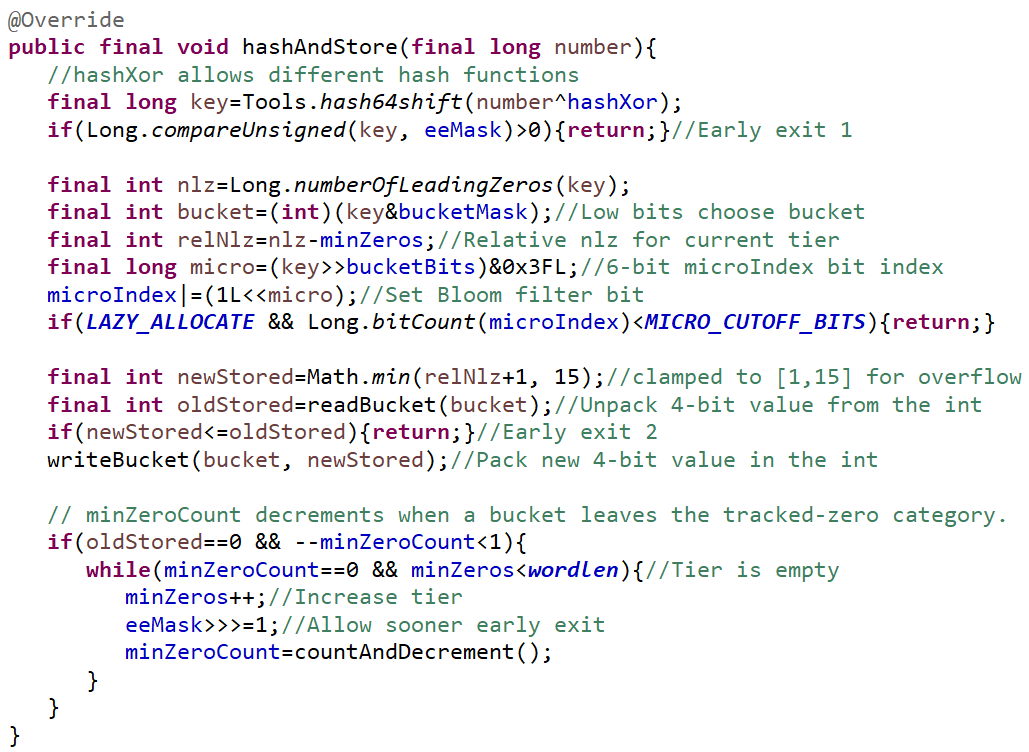}
\caption{\textbf{Listing 1.} The complete \texttt{hashAndStore} method
for DLL4. The early exit mask (line 4) rejects the vast majority of
elements at high cardinality before any bucket access. MicroIndex (lines
7--8) provides a Bloom-filter floor for low cardinality. The tier
promotion loop (lines 16--21) advances \texttt{minZeros} and tightens
the early exit mask when all buckets have been filled.}
\end{figure}

\begin{figure}
\centering
\includegraphics{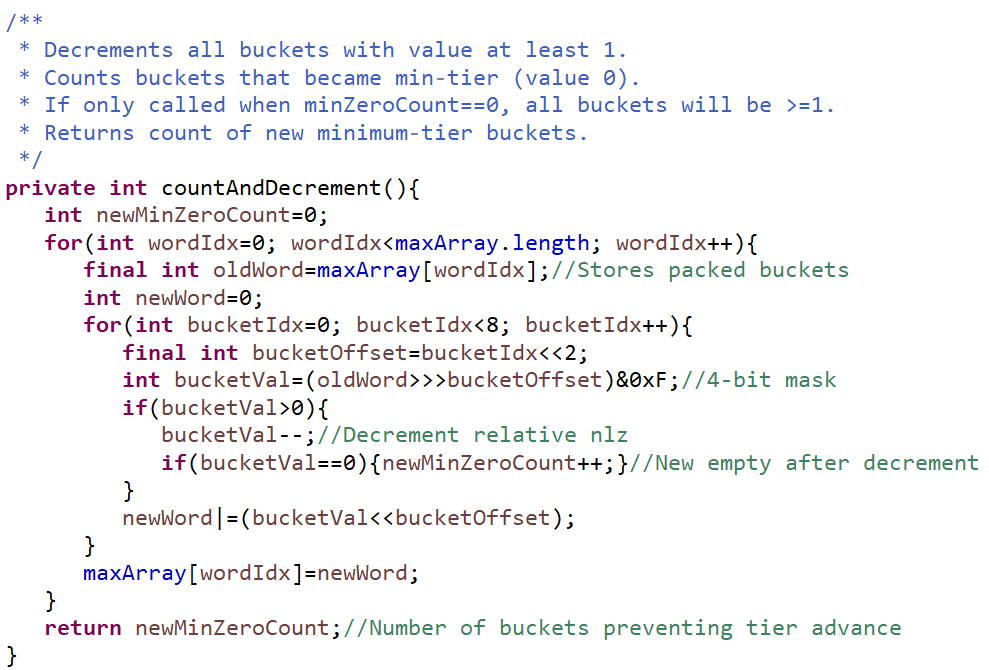}
\caption{\textbf{Listing 2.} The \texttt{countAndDecrement} method,
called during tier promotion. Decrements every non-empty bucket's stored
value by 1 and counts how many reach the minimum tier, determining
whether another promotion is needed.}
\end{figure}

\hypertarget{memory-comparison}{%
\subsection{3.4 Memory Comparison}\label{memory-comparison}}

\begin{longtable}[]{@{}lllll@{}}
\toprule
Variant & Bits/bucket & 2048 buckets & 4096 buckets & Relative to
HLL-2048\tabularnewline
\midrule
\endhead
HLL (6-bit) & 6 & 1,536 B & 3,072 B & 1.00\(\times\)\tabularnewline
DLL4 (4-bit) & 4 & 1,024 B & 2,048 B & 0.67\(\times\)\tabularnewline
DLL3 (3-bit) & 3 & 768 B & 1,536 B & 0.50\(\times\)\tabularnewline
\bottomrule
\end{longtable}

At equal memory, DLL4 uses 50\% more buckets than HLL, reducing the
standard error by a factor of \(\sqrt{3/2}\) \(\approx\) 1.22. DLL3 uses
double the buckets, reducing error by \(\sqrt{2}\) \(\approx\) 1.41 ---
though with some limitations for low-complexity data (Section 5.3).

DLL4's 4-bit packing is additionally hardware-friendly: 8 buckets per
32-bit word with zero wasted bits. HLL's 6-bit packing wastes 2 bits per
32-bit word (only 5 buckets per word) or requires cross-word spanning.

Beyond the bucket array, DLL requires a small constant overhead: one
byte for the current tier floor (\texttt{minZeros}), which is the only
architecturally required auxiliary state to allow cardinality over
2\^{}255. Our implementation additionally stores a 64-bit MicroIndex for
improved low-cardinality estimation and cached early-exit state (eeMask
+ counter) for speed, totaling approximately 28 bytes of overhead
independent of bucket count --- negligible relative to the bucket array
at any practical size.

\hypertarget{dynamic-linear-counting}{%
\section{4. Dynamic Linear Counting}\label{dynamic-linear-counting}}

\hypertarget{the-lc-ceiling-problem}{%
\subsection{4.1 The LC Ceiling Problem}\label{the-lc-ceiling-problem}}

Linear Counting estimates cardinality from the fraction of empty
buckets:

\[\text{LC} = B \cdot \ln\!\left(\frac{B}{V}\right)\]

This is remarkably accurate at low cardinality --- when \emph{V} is
large relative to \emph{B}. But LC has a hard ceiling: once all buckets
are filled (\emph{V} = 0), the estimate becomes infinite. In practice,
LC degrades rapidly as \emph{V} shrinks; above \textasciitilde5\emph{B},
its variance is too large for reliable use as a blending component.

HyperLogLog addresses this by switching to the harmonic mean estimator
above a threshold. But the transition is the source of HLL's
characteristic error spike: in the crossover region, neither estimator
is well-suited, and the discontinuity between them introduces additional
error.

\hypertarget{dlc-tier-aware-linear-counting}{%
\subsection{4.2 DLC: Tier-Aware Linear
Counting}\label{dlc-tier-aware-linear-counting}}

DLL's tier structure offers a natural solution. At any tier \emph{t},
define the \emph{tier-t empty count}:

\[V_t = V + \sum_{i=0}^{t-1} n_i\]

where \emph{V} is the number of truly empty buckets and \emph{n\_i} is
the number of buckets with absolute NLZ equal to \emph{i}. Conceptually,
\emph{V\_t} treats all buckets with NLZ below \emph{t} as ``empty for
the purposes of tier \emph{t}.''

The Dynamic Linear Counting estimate at tier \emph{t} is then:

\[\text{DLC}(t) = 2^t \cdot B \cdot \ln\!\left(\frac{B}{V_t}\right)\]

At tier 0, this reduces to classic LC. At tier 1, buckets with NLZ = 0
are added to the ``empty'' pool, doubling the effective range. Each
successive tier extends the range by another factor of 2. Where LC is
accurate up to \textasciitilde2.5\emph{B}, DLC(\emph{t}) is accurate up
to approximately 2\^{}\emph{t} \(\times\) 2.5\emph{B}.

The key insight is that \textbf{DLC provides a useful estimate at every
cardinality}, not just at low cardinality. At any given cardinality,
there exists some tier \emph{t} where \emph{V\_t} is near the optimal
range (roughly \emph{B}/6 to \emph{B}/3, centered on the optimal
\emph{V} \(\approx\) \emph{B}/4 where LC's estimation error is
minimized). DLC at that tier gives a low-error estimate (1.617\% mean
absolute error without CF; see Section 8.3, Table 3 and Figure 6)
without needing the harmonic mean, correction factors, or a transition
function. DLL's Hybrid blend (Section 4.6) still uses empirically chosen
boundaries (0.2\emph{B} and 5\emph{B}), similar to HLL's 2.5\emph{B}
threshold --- but the critical difference is that both estimators in
DLL's blend (LCmin and Mean) are accurate throughout the crossover zone,
so the exact cutoff values are not load-bearing. HLL's threshold is
fragile because it switches between a failing estimator (LC) and a
not-yet-accurate one (harmonic mean); DLL's boundaries merely define a
window over which two good estimates are smoothly combined.

\hypertarget{lcmin-tier-compensated-linear-counting}{%
\subsection{4.3 LCmin: Tier-Compensated Linear
Counting}\label{lcmin-tier-compensated-linear-counting}}

The simplest DLC variant uses the tier floor directly:

\[\text{LCmin} = 2^{\text{minZeros}} \cdot B \cdot \ln\!\left(\frac{B}{V}\right)\]

This equals DLC at the lowest active tier (the tier corresponding to
\texttt{minZeros}). After each tier promotion, the empty-bucket count
resets and LC becomes accurate again --- scaled by the accumulated
promotion factor. LCmin serves as the low-cardinality anchor in DLL's
hybrid blend.

\hypertarget{dlcbest-best-single-tier-estimate}{%
\subsection{4.4 DLCbest: Best Single-Tier
Estimate}\label{dlcbest-best-single-tier-estimate}}

For each tier \emph{t}, compute \emph{V\_t} and select the tier whose
\emph{V\_t} is closest to a target occupancy (empirically, 25\% free ---
i.e., \emph{V\_t} \(\approx\) \emph{B}/4):

\[t^* = \arg\min_t |V_t - B/4|\]

DLCbest returns DLC(t*). When two adjacent tiers are equidistant from
the target, their estimates are averaged. This produces the most
accurate single-point estimate at any cardinality, though it can be
slightly noisier than the blended DLC due to tier switching.

\hypertarget{dlc-variance-weighted-log-space-blend}{%
\subsection{4.5 DLC: Variance-Weighted Log-Space
Blend}\label{dlc-variance-weighted-log-space-blend}}

The production DLC estimator weights each tier by the inverse of its
theoretical mean absolute error. At tier \emph{t}, the LC estimate uses
\(V_t\) empty buckets out of \emph{B} total. Since \(V_t\) follows a
binomial distribution with variance \(V_t(B - V_t)/B\), the delta method
gives the standard error of \(\text{DLC}(t)\) as:

\[\sigma_t = \frac{\sqrt{(B - V_t)/(B \cdot V_t)}}{\ln(B/V_t)}\]

Converting to expected mean absolute error (for a normal distribution,
\(E[|X|] = \sqrt{2/\pi} \cdot \sigma\)):

\[\text{MAE}_t = \sqrt{\frac{2}{\pi}} \cdot \sigma_t\]

The weight for each tier is the inverse MAE raised to a sharpening power
\(p = 4.5\):

\[w_t = \left(\frac{1}{\text{MAE}_t}\right)^p\]

The final DLC estimate blends all valid tiers in log-space:

\[\text{DLC} = \exp\!\left(\frac{\sum_t w_t \cdot \ln(\text{DLC}(t))}{\sum_t w_t}\right)\]

Tiers near the optimal occupancy (where \(V_t \approx B/4\)) naturally
receive the highest weight because their LC estimates have the lowest
variance. Tiers with very few or very many empties --- where LC is
either saturated or starved of information --- receive negligible
weight. At very low cardinality (\(V > 0.3B\)), DLC transitions smoothly
to LCmin because few tiers are informative.

The only fitted parameter is the sharpening power \(p = 4.5\), which
controls how aggressively low-variance tiers dominate the blend. The
optimum is broad: values from 3 to 6 produce similar accuracy, with 4.5
minimizing error in simulation. The weight function itself requires no
calibration because it is derived from the theoretical variance of
Linear Counting. This makes DLC a nearly parameter-free estimator.

DLC achieves accuracy comparable to CF-corrected HLL across the full
cardinality range --- without requiring a correction factor table. This
is unique among the estimators evaluated: HLL, Mean, and Hybrid all
require CF tables for optimal accuracy, and even FGRA (Section 8.3)
embeds correction constants (\(\sigma\)/\(\varphi\) series) derived from
theoretical analysis. DLC is the only estimator that covers the full
cardinality range with no pre-computed corrections of any kind. When
correction factors \emph{are} applied, DLC's accuracy improves further,
as the CF addresses the small systematic biases inherent in the LC
formula.

\hypertarget{logarithmic-hybrid-blend-eliminating-the-bulge}{%
\subsection{4.6 Logarithmic Hybrid Blend: Eliminating the
Bulge}\label{logarithmic-hybrid-blend-eliminating-the-bulge}}

DLL's hybrid estimator --- the \emph{Logarithmic Hybrid Blend} ---
smoothly transitions between LCmin and the CF-corrected Mean estimator
using a linearly interpolated blend with a logarithmically scaled mixing
weight:

\[\text{Hybrid} = \begin{cases}
\text{LCmin} & \text{if LCmin} \leq 0.2B \\
(1-t) \cdot \text{LCmin} + t \cdot \text{Mean}_{\text{CF}} & \text{if } 0.2B < \text{LCmin} < 5B \\
\text{Mean}_{\text{CF}} & \text{if LCmin} \geq 5B
\end{cases}\]

where \emph{t} = ln(LCmin / 0.2\emph{B}) / ln(5\emph{B} / 0.2\emph{B}).
The mixing weight \emph{t} is logarithmic in LCmin, providing smooth
transitions at both endpoints.

The critical differences from HLL's transition are that both endpoints
are accurate in the crossover region, and the entire transition region
is more accurate than either input curve, whereas HLL is less accurate
than either input, and usually both. LCmin, being tier-compensated,
remains accurate much further into the transition zone than raw LC. The
Mean estimator with correction factor is accurate from moderate
cardinality upward. The blend smoothly combines two good estimates,
rather than switching between a failing estimate (LC) and a
not-yet-accurate one (harmonic mean), as in HyperLogLog.

The result is a hybrid estimator with no error spike: the crossover
region is as flat as the regions on either side of it. The blend region
exceeds the accuracy of either input because the two estimates often
exhibit error in opposite directions, resulting in partial cancellation.

\begin{figure}
\centering
\includegraphics{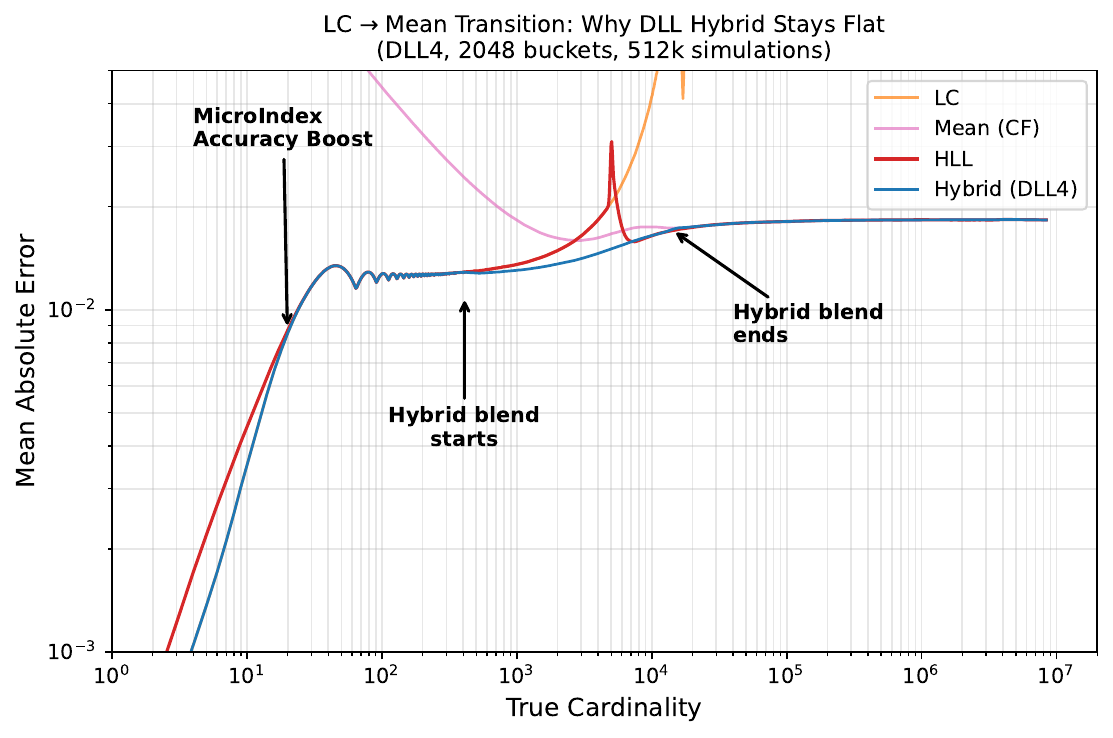}
\caption{\textbf{Figure 2.} Absolute error of LC, Mean (CF), HLL, and
DLL Hybrid as cardinality increases, on a log Y-axis. LC and HLL diverge
in the transition region while the Hybrid remains flat throughout. The
reduced error of Hybrid compared to HLL in the 400--10,500 cardinality
region is a combination of the improved blend between LC and Mean, and
the inherent lower error of Mean compared to HMean (see Figures 4--5).
Below cardinality \textasciitilde20, the MicroIndex provides additional
accuracy. (DLL4, 2048 buckets, 512k simulations.)}
\end{figure}

\hypertarget{correction-factors}{%
\section{5. Correction Factors}\label{correction-factors}}

\hypertarget{cf-table-structure}{%
\subsection{5.1 CF Table Structure}\label{cf-table-structure}}

Like HyperLogLog, DLL benefits from correction factors (CFs) that
compensate for systematic biases in the raw estimators. DLL uses a
cardinality-indexed CF table: the correction factor for each estimator
type is stored as a function of true cardinality.

The CF is applied as a single multiplicative correction: the raw
estimate is used as the lookup key into the CF table, and the corrected
estimate is the raw estimate times the CF value at that cardinality.
Because the CF curves are nearly flat (close to 1.0 throughout), a
single lookup suffices without iteration.

CF tables are generated by simulation: many independent estimator
instances with different hash functions are run across a range of known
cardinalities, and the average bias at each cardinality point yields the
multiplicative correction factor.

\begin{figure}
\centering
\includegraphics{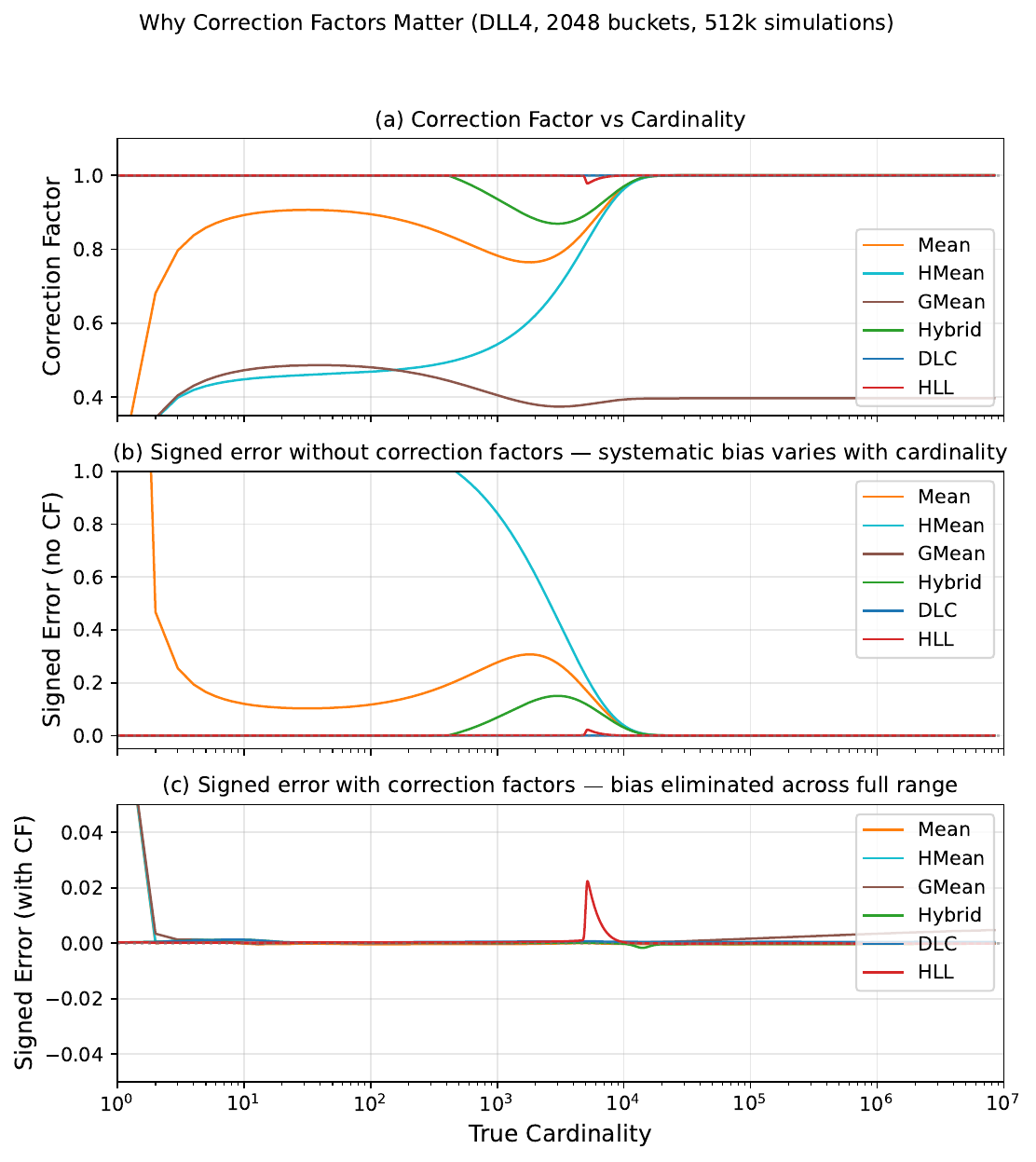}
\caption{\textbf{Figure 3.} Three-panel demonstration of correction
factors. (a) Implied CF values by estimator type. (b) Signed error
without CF --- large systematic biases are visible. (c) Signed error
with CF --- bias is eliminated across the full cardinality range. (DLL4,
2048 buckets, 512k simulations.)}
\end{figure}

\hypertarget{self-similar-cf-lookup}{%
\subsection{5.2 Self-Similar CF Lookup}\label{self-similar-cf-lookup}}

In many cases, such as DLL4 Hybrid, the CF appears to asymptotically
approach a constant, but tier promotion effects are more pronounced for
DLL3 due to a higher overflow rate, or DLC using any estimator type.
DLL's tier structure is self-similar: the bucket distribution after each
tier promotion is statistically identical to the previous era, scaled by
a factor of 2. This means the correction factor pattern repeats every
cardinality doubling:

\[\text{CF}(2C) \approx \text{CF}(C)\]

The CF table is generated at a fixed range --- typically 2,048 buckets
\(\times\) 4,096 maximum multiplier \(\approx\) 8,388,608 maximum
cardinality, with 1\% logarithmic key spacing yielding approximately
1,600 entries per estimator type. To look up a correction factor at any
cardinality beyond the table's range, the estimate is simply
right-shifted until it falls within bounds:

\begin{verbatim}
while (estimate > tableMax) estimate >>>= 1;
CF = table.lookup(estimate);
\end{verbatim}

Because the CF pattern repeats every cardinality doubling, this produces
the correct correction factor regardless of the actual cardinality. The
table is therefore a small, fixed-size constant --- approximately 1,600
entries per estimator type --- that serves the full cardinality range
from 1 to 2\^{}63 without extension or tiling. Only one CF table is
needed regardless of the number of estimator instances.

\hypertarget{bucket-averaging-methods}{%
\subsection{5.3 Bucket Averaging
Methods}\label{bucket-averaging-methods}}

All LogLog-family estimators must combine the maximum NLZ values stored
across \emph{B} buckets into a single cardinality estimate. Several
averaging methods are possible; we evaluate three.

Let \emph{NLZ\_j} denote the absolute NLZ stored in bucket \emph{j}, and
let the sum run over filled buckets only (\emph{count} = number of
filled buckets). Define the \emph{difference value} \emph{d\_j} =
2\^{}(63 \(-\) \emph{NLZ\_j}), which is proportional to the probability
of observing NLZ \(\geq\) \emph{NLZ\_j} in a uniform 64-bit hash.

\textbf{Occupancy-corrected Mean.} Computes the arithmetic mean of the
difference values \(d_j = 2^{63 - NLZ_j}\) and inverts:

\[\text{Mean} = \frac{2^{64} \cdot \text{count}^2 \cdot (\text{count} + B)}{2B \cdot \sum_j d_j}\]

Substituting \(d_j = 2^{63} \cdot 2^{-NLZ_j}\) and simplifying reveals:

\[\text{Mean} = \frac{\text{count} + B}{2B} \cdot \frac{2 \cdot \text{count}^2}{\sum_j 2^{-NLZ_j}}\]

\textbf{HyperLogLog's harmonic mean (HMean).} The classic HLL formula
applied over filled buckets only:

\[\text{HMean} = \alpha_m \cdot \frac{\text{count}^2}{\sum_j 2^{-NLZ_j}}\]

where \(\alpha_m = 0.7213/(1 + 1.079/B)\).

\textbf{Both formulas are fundamentally harmonic means} of the
underlying bucket probabilities \(2^{-NLZ_j}\), sharing the same core:
\(2 \cdot \text{count}^2 / \sum_j 2^{-NLZ_j}\). The \emph{only}
difference is the leading coefficient. HMean uses Flajolet's static
constant \(\alpha_m\), which is derived under the assumption that all
buckets are occupied. Mean uses the dynamic, occupancy-aware coefficient
\((\text{count} + B)/(2B)\), which tracks the actual number of filled
buckets.

This distinction explains three empirically observed regimes:

\begin{enumerate}
\def\labelenumi{\arabic{enumi}.}
\item
  \textbf{High cardinality (\(> 7.02 \times B\)):} All buckets are
  filled (\(\text{count} = B\)), so Mean's coefficient reduces to
  \((B + B)/(2B) = 1\) --- a fixed constant, comparable to
  \(\alpha_m \approx 0.72\). Because both formulas are now static scalar
  multiples of \(2 \cdot \text{count}^2 / \sum_j 2^{-NLZ_j}\), the CF
  table maps them to identical values with identical variance.
\item
  \textbf{Low cardinality (\(< 0.71 \times B\)):} The vast majority of
  buckets are empty, and non-empty buckets almost all have
  \(NLZ_j = 1\). Both formulas effectively degenerate to tracking the
  occupied count, rendering them equivalent.
\item
  \textbf{Transition zone (\(0.71 \times B\) to \(7.02 \times B\)):}
  Hash collisions cause multiple elements to land in the same bucket, so
  the number of filled buckets (\(\text{count}\)) is less than the true
  cardinality. Mean's dynamic coefficient \((\text{count} + B)/(2B)\)
  compensates for these collisions by adjusting the estimate based on
  actual occupancy. HMean's static \(\alpha_m\) cannot account for this.
\end{enumerate}

In other words, Mean is not a fundamentally different estimator from
HMean --- it is a \emph{generalization} that replaces Flajolet's
constant-occupancy approximation with an exact occupancy correction.
HMean is the special case of Mean where the coefficient is frozen. Mean
is never worse than HMean at any cardinality tested, and is modestly but
consistently better in the transition zone (e.g., 1.613\% vs 1.645\%
mean absolute error at \(1 \times B\)).

\textbf{Geometric Mean (GMean).} Computes the geometric mean of the
difference values and inverts:

\[\text{GMean} = 2 \cdot \frac{2^{63}}{\exp\!\left(\frac{1}{\text{count}} \sum_j \ln(d_j)\right)} \cdot \frac{\text{count} + B}{2B}\]

By the AM--GM inequality, the geometric mean of \(d_j\) is always
\(\leq\) the arithmetic mean, so GMean produces a larger raw estimate
than Mean. This translates to a large positive bias before correction.
After CF correction, GMean achieves roughly twice the absolute error of
Mean and HMean throughout the cardinality range. CF correction removes
bias but cannot compensate for GMean's higher intrinsic variance.

We use Mean in DLL's hybrid estimator because it is provably at least as
accurate as HMean and requires no special constant.

\begin{figure}
\centering
\includegraphics{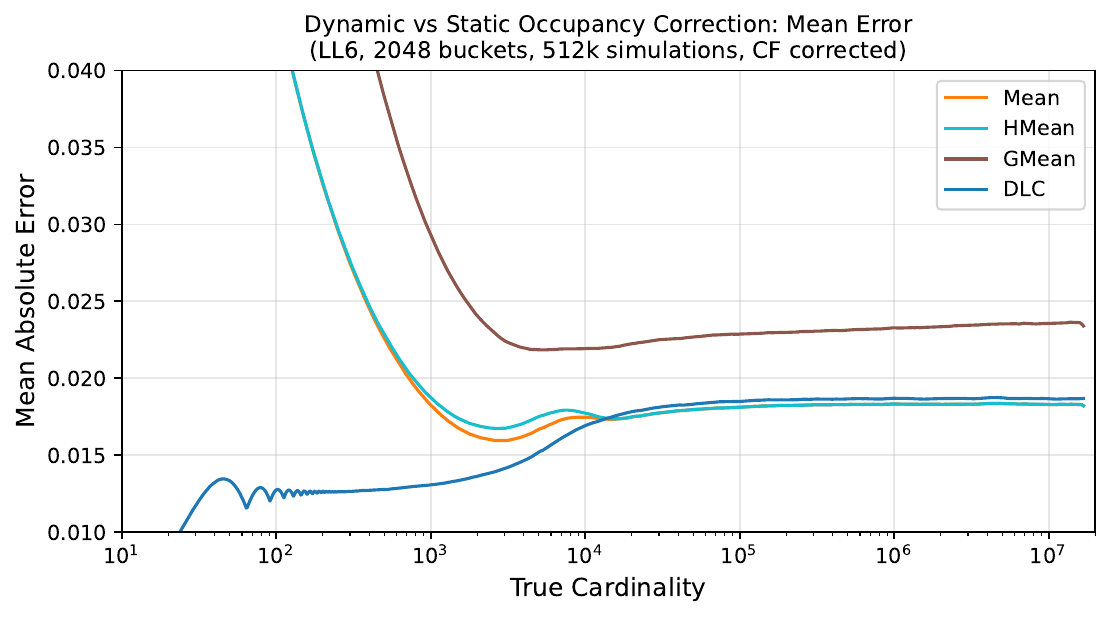}
\caption{\textbf{Figure 4.} Mean absolute error after CF correction for
Mean, HMean, GMean, and DLC. Mean and HMean are both harmonic means of
the same bucket probabilities; the only difference is Mean's dynamic
occupancy coefficient (count+B)/(2B) versus HMean's static
\(\alpha\)\_m. Below 0.71\(\times\)B and above 7.02\(\times\)B (arrows),
occupancy is effectively constant and the two converge. In the
transition region, Mean's dynamic coefficient absorbs occupancy
fluctuations, yielding modestly lower error. GMean is consistently
worse. (LL6, 2048 buckets, 512k simulations.)}
\end{figure}

\begin{figure}
\centering
\includegraphics{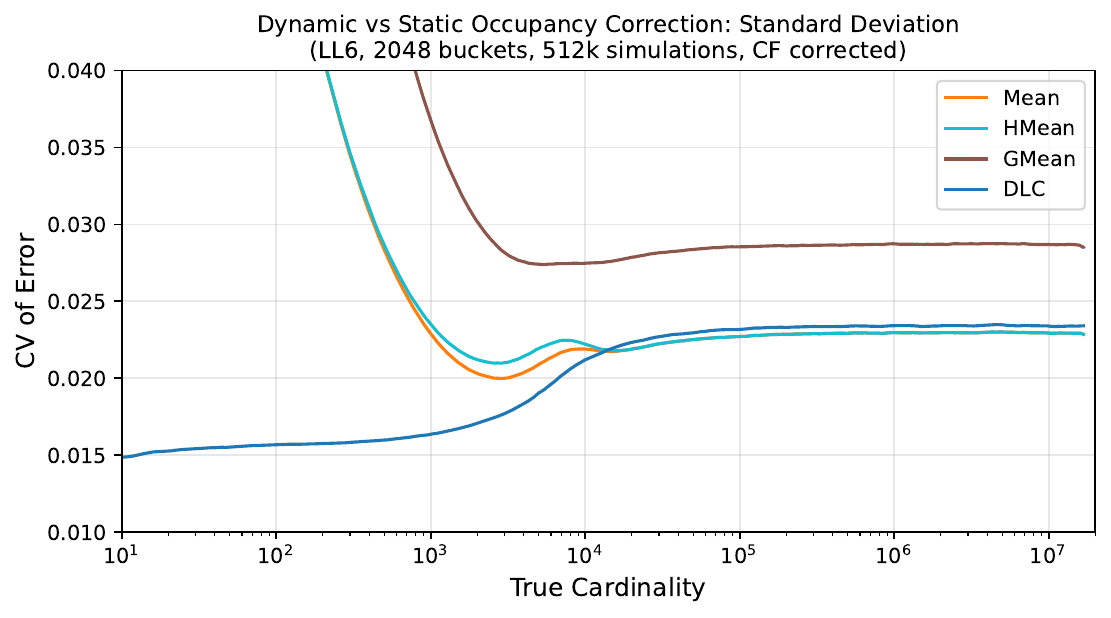}
\caption{\textbf{Figure 5.} Standard deviation of error after CF
correction for the same estimators. Mean and HMean show identical
standard deviation outside the 0.71\(\times\)--7.02\(\times\) bucket
range, confirming that when occupancy is constant, the dynamic and
static coefficients produce identical variance. The difference is
confined to the partial-occupancy region where collisions are common.
(LL6, 2048 buckets, 512k simulations.)}
\end{figure}

\hypertarget{dll3-3-bit-variant-and-overflow-correction}{%
\subsection{5.4 DLL3: 3-Bit Variant and Overflow
Correction}\label{dll3-3-bit-variant-and-overflow-correction}}

DLL3 uses 3 bits per bucket (stored values 0--7), encoding
\texttt{stored\ =\ relNlz}, yielding relative NLZ values 0 through 7.
This reduces memory by 25\% compared to DLL4 --- or equivalently, allows
33\% more buckets at the same memory.

The tradeoff is that buckets whose relative NLZ exceeds 6 are clamped to
stored = 7 (the maximum), creating a systematic underestimate at high
cardinality. No stored value is reserved for overflow; stored = 7
represents both legitimate relNlz = 6 and clamped overflow. Each
affected tier accumulates approximately
\emph{B}\(\cdot\)ln(2\emph{B})/256 ``ghost'' overflow values per era.

\textbf{Recording overflow.} At each tier promotion, DLL3 counts the
number of buckets at stored = 7 (the maximum) and records half this
count as the estimated overflow for the next tier:

\[\text{storedOverflow}[t] = \lfloor\text{topCount} / 2\rfloor\]

The factor of 1/2 arises from the geometric distribution: half the
buckets at stored = 7 have true NLZ exactly equal to 6 + minZeros, while
the other half have NLZ strictly greater (overflow victims).

\textbf{Cumulative-space correction.} The correction operates in
reverse-cumulative space rather than on individual tier counts:

\[\text{cumRaw}[t] = \sum_{i=t}^{63} n_i\]

For each affected tier:

\[\text{corrCum}[t] = \text{cumRaw}[t] + (B - \text{cumRaw}[t]) \cdot (1 - e^{-X_t/B})\]

where X\_t is the stored overflow estimate for tier t. Working in
cumulative space is essential: it makes each tier's correction
independent, avoiding the cascading errors that arise when correcting
individual tier counts. The corrected single-tier counts are recovered
as differences of adjacent cumulative values.

With this correction, DLL3 matches DLL4's accuracy within 0.02\% for
high-complexity (all-unique) data. However, for low-complexity data with
many duplicate values, tier promotions caused by duplicates interact
with the overflow correction to produce a systematic positive bias above
\textasciitilde100\(\times\)\emph{B} cardinality. DLL4, with its 16-tier
range, is unaffected (see Figure 9).

\textbf{Recommendation:} DLL4 (4-bit) is the recommended default. DLL3
(3-bit) outperforms DLL4 in the same memory space only when the stream
is known a priori to have near-maximal complexity.

\hypertarget{microindex}{%
\section{6. MicroIndex}\label{microindex}}

For very small cardinalities (below \textasciitilde120 distinct
elements), a 64-bit Bloom filter {[}15{]} is sufficient for accurate
cardinality estimates, without allocating a bucket array. DLL addresses
this with a \emph{MicroIndex}: a single 64-bit integer that can be used
to either lazy-allocate the bucket array once cardinality exceeds a
threshold (resulting in larger cardinalities being underestimated by up
to that threshold), or to increase the accuracy of LC in the
low-cardinality region by adding hash-collision resilience.

Six bits from the hash value (above the bucket selector bits) index into
the 64-bit word, setting the corresponding bit. The resulting population
count gives a miniature Linear Counting estimate over 64 virtual
``buckets'':

\[\text{MicroEst} = 64 \cdot \ln\!\left(\frac{64}{\max(64 - \text{popcount}(\text{microIndex}), 0.5)}\right)\]

This is used as the best estimate prior to lazy array allocation. When
the bucket array is allocated, MicroIndex provides a cardinality floor:
at low cardinality, multiple elements can hash to the same main bucket
but different MicroIndex positions, so the MicroIndex population count
may exceed the number of filled buckets. The LC estimate is floored by
this count:

\[\text{LC} = \max\!\big(\text{lc}(B, V),\; \text{popcount}(\text{microIndex}),\; \text{historyFloor}\big)\]

where historyFloor is the number of filled buckets plus the total set
history bits. This ensures that the estimate never falls below the
number of distinct hash patterns observed by the MicroIndex, improving
accuracy for cardinalities below \textasciitilde64 where hash collisions
within the main bucket array are still common.

The MicroIndex also enables lazy array allocation: the main bucket array
need not be created until the MicroIndex saturates, saving memory for
applications that create many estimators (e.g., per-\emph{k}-mer
tracking in bioinformatics). The MicroIndex requires zero additional
memory beyond a single \texttt{long} field.

Additionally, in practice, cardinality estimates are capped at the
number of elements added (which eliminates overestimation in
maximally-complex streams), though this clamping was disabled for all
figures in this paper.

\hypertarget{methods}{%
\section{7. Methods}\label{methods}}

\hypertarget{simulation-framework}{%
\subsection{7.1 Simulation Framework}\label{simulation-framework}}

All accuracy evaluations used purpose-built calibration drivers included
in BBTools. Two simulators were developed to test estimator accuracy
under different data distributions:

\textbf{High-complexity simulator (DDLCalibrationDriver2).} This driver
tests accuracy on streams of all-unique elements. Each thread creates
one estimator instance at a time, feeds it uniformly random 64-bit
values from a seeded PRNG (Xoshiro256++ {[}16{]}), hashed through Thomas
Wang's 64-bit integer hash function {[}14{]}, records estimator output
at logarithmically-spaced cardinality thresholds (1\% increments), then
discards the instance and creates the next. This one-at-a-time design
keeps the working set in L1/L2 cache for the entire run, eliminating
cache effects from the accuracy measurement. The driver is fully
deterministic: the same master seed produces identical results
regardless of thread count. Results across all instances are merged
after all threads complete.

For each threshold, the driver records the signed relative error
(estimate \(-\) true)/true, the absolute relative error, and the squared
error for standard deviation computation. Correction factor tables are
generated from the same output by computing CF = true/estimate averaged
across all instances at each cardinality point.

\textbf{Low-complexity simulator (LowComplexityCalibrationDriver).} This
driver tests accuracy on streams with bounded cardinality and many
duplicate elements, simulating real-world data where the same elements
appear repeatedly. Rather than materializing a full array of
\texttt{cardinality} 64-bit values (which would thrash cache at high
cardinalities), two int arrays of length \(\sqrt{}\)cardinality are
generated from a master seed; virtual values are composed on the fly as
Cartesian products of lower and upper 32-bit halves. Each estimator
instance draws from this virtual array with replacement, biased toward
lower indices via min(rand(), rand()) to produce a skewed frequency
distribution resembling natural data.

Elements are added via \texttt{hashAndStore()} on every draw (including
duplicates). True cardinality is tracked via a BitSet recording which
unique values have been seen. Estimates are recorded on every add while
the true cardinality is at a reporting threshold, capturing how the
estimator's state evolves as duplicate values trigger tier promotions.
Between thresholds, only \texttt{hashAndStore()} runs --- no
\texttt{rawEstimates()} overhead.

The \texttt{iterations} parameter controls how long each estimator runs
beyond saturation: \texttt{iterations=N} means each estimator processes
cardinality\(\times\)N total adds. This allows testing behavior at
extreme duplication levels where the ratio of total adds to unique
elements is very high.

\textbf{Reporting.} Both simulators report results at
exponentially-spaced cardinality thresholds: each threshold is at least
1\% larger than the previous, yielding approximately 230 reporting
points per decade of cardinality. Error metrics are the mean across all
estimator instances at each threshold: signed error (bias), absolute
error (accuracy), and standard deviation (precision).

\hypertarget{baseline-comparator}{%
\subsection{7.2 Baseline Comparator}\label{baseline-comparator}}

For fair comparison, we implemented LL6 (LogLog6), a 6-bit-per-bucket
estimator that serves as a structural equivalent of HyperLogLog. LL6
stores one byte per bucket (6 bits used, 2 wasted), has no tier
promotion, no shared exponent, and no eeMask early exit. Every hash
accesses a bucket. LL6 uses the same estimator pipeline as DLL4 --- the
same \texttt{CardStats} path, the same DLC computation, the same hybrid
blend --- isolating the effect of DLL's architectural innovations
(shared exponent, tier promotion, eeMask) from estimator algorithm
differences.

LL6's correction factor table was generated independently using the
high-complexity simulator with 512,000 instances at 2,048 buckets.

Note that LL6 is a structural baseline, not a reimplementation of
HyperLogLog++ {[}4{]}. HLL++ adds empirical bias correction tables
(k-nearest-neighbor interpolation at 200 calibration points) and a
sparse representation at higher virtual precision for small
cardinalities, which partially reduce the transition spike but do not
eliminate it. Our LL6 baseline uses DLL's estimator pipeline (including
DLC and Hybrid), isolating the comparison to architectural differences
rather than estimation algorithm differences.

\hypertarget{speed-benchmarks}{%
\subsection{7.3 Speed Benchmarks}\label{speed-benchmarks}}

Speed benchmarks were conducted on two compute nodes: (1) a 64-core node
(AMD EPYC 7502P, 2.5 GHz; 32 KB L1d, 512 KB L2 per core; 128 MB shared
L3; 512 GB RAM) for the BBDuk benchmark (Figure 7) and cache thrashing
benchmark (Figure 8); and (2) a 128-core node (same architecture,
dual-socket) for the per-type speed comparison (Table 7). The BBDuk
benchmark processed 325 million 150bp Illumina reads (48.8 Gbp) with 16
threads, measuring end-to-end throughput in Mbp/s. The cache thrashing
benchmark used DDLCalibrationDriver in single-threaded benchmark mode
(\texttt{benchmark=t}), running pure \texttt{hashAndStore()} loops with
no estimate computation. Table 7 speed values were measured with 128
threads and divided by 128 to yield per-core throughput.

\hypertarget{experimental-results}{%
\section{8. Experimental Results}\label{experimental-results}}

\hypertarget{high-complexity-evaluation}{%
\subsection{8.1 High-Complexity
Evaluation}\label{high-complexity-evaluation}}

We simulated 512,000 independent estimator instances with 2,048 buckets
each, out to a true cardinality of 8,388,608 (4,096\(\times\) the bucket
count), sampled at 1,297 exponentially-spaced checkpoints.

\textbf{The HLL transition bump (Figure 1).} HLL exhibits an error spike
in the transition region between Linear Counting and the harmonic mean
estimator, peaking at 3.1\% mean absolute error near cardinality 5,000
(2.5\(\times\) \emph{B}) --- approximately 70\% above steady-state error
levels. In contrast, DLL4's Hybrid estimator and DLC both remain below
2\% error throughout the entire range, with no visible transition
artifact.

\textbf{Stable estimation methods.} Hybrid (with CF) achieves 1.572\%
log-weighted average error; DLC (without CF) achieves 1.617\%; HLL
achieves 1.607\%. Log-weighted averages are similar because HLL's
transition bump is narrow; the difference is visible primarily in peak
error: Hybrid peaks at 1.835\% vs HLL's 3.101\%. FGRA also eliminates
the bump entirely via its single closed-form estimator (see Section
8.3).

\begin{figure}
\centering
\includegraphics{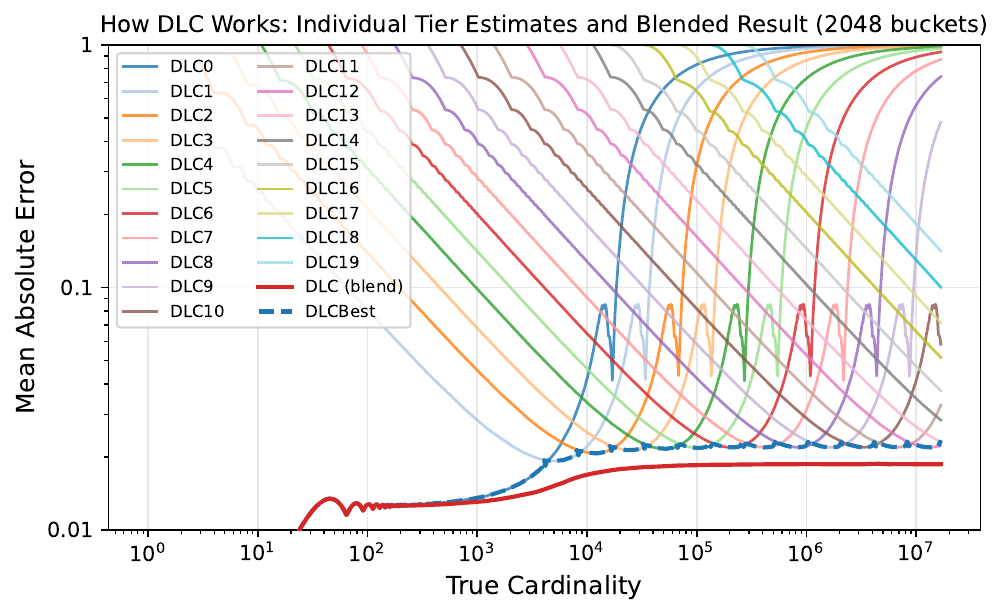}
\caption{\textbf{Figure 6.} Individual DLC tier estimates (DLC0--DLC19)
each have a U-shaped accuracy curve accurate near their optimal
cardinality range. The DLC blend (red) threads through the valleys of
all tier curves. (DLL4, 2048 buckets, 512k simulations.)}
\end{figure}

\textbf{DLC tier structure (Figure 6).} The individual DLC tier
estimates (DLC0 through DLC19) each have a U-shaped accuracy curve:
accurate near their optimal cardinality range and diverging above and
below. The tiers tile the cardinality range like overlapping scales,
each covering approximately one octave. The DLC blend (red) threads
through the valleys of all tier curves, selecting the optimal tier at
each cardinality.

\begin{figure}
\centering
\includegraphics{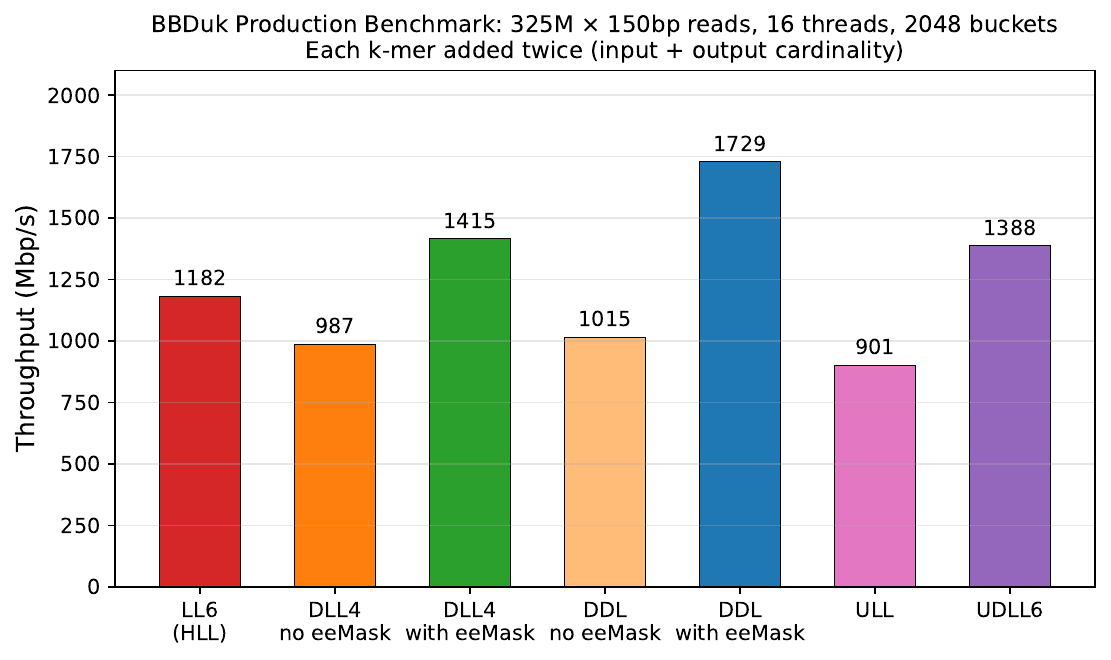}
\caption{\textbf{Figure 7.} BBDuk throughput (Mbp/s) processing 325M
reads with 16 threads, 2,048 buckets. As a QC tool, BBDuk measures k-mer
cardinality for both input and output reads (before and after
filtering/trimming), so each k-mer is added twice, using two estimators
per thread. DLL4 with eeMask achieves 1,415 Mbp/s; without eeMask it
drops to 987 Mbp/s. DDL achieves 1,729 Mbp/s. ULL achieves 901 Mbp/s;
UDLL6 achieves 1,388 Mbp/s (54\% faster than ULL).}
\end{figure}

\textbf{Speed (Figure 7).} In a production BBDuk run processing 325
million 150bp reads with 16 threads, DLL4 achieved 1,415 Mbp/s compared
to 1,182 Mbp/s for LL6 (20\% faster). Disabling eeMask reduced DLL4 to
987 Mbp/s (17\% slower than LL6), demonstrating that eeMask is essential
to overcome tier promotion overhead. DDL, which uses the same eeMask
with 8-bit buckets and simpler byte-array packing, achieved 1,729 Mbp/s
(70\% faster than without eeMask). ULL achieved 901 Mbp/s and UDLL6
achieved 1,388 Mbp/s (54\% faster than ULL), nearly matching DLL4. ULL
lacks early exit entirely; UDLL6's int-packed registers and conservative
eeMask recover most of the speed (see Section 8.3 for details).

\begin{figure}
\centering
\includegraphics{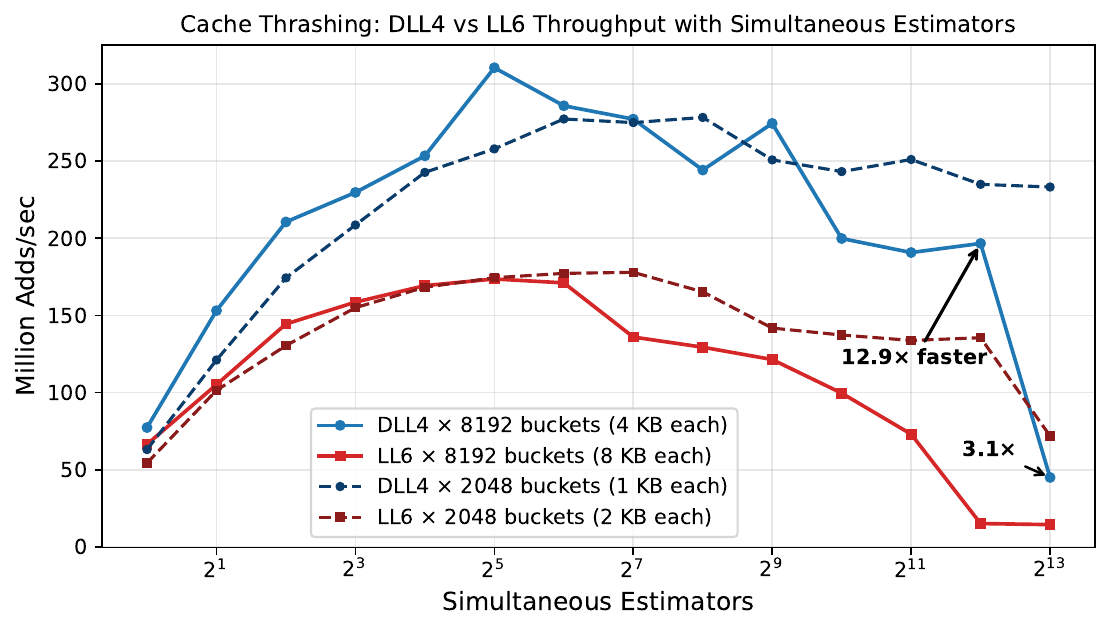}
\caption{\textbf{Figure 8.} Throughput (million adds/sec,
single-threaded) vs number of simultaneous estimator instances, for DLL4
and LL6 at 2048 and 8192 buckets. LL6's larger per-instance footprint
causes severe cache thrashing at high counts; DLL4's eeMask eliminates
most memory accesses regardless of cache pressure. At 4096 simultaneous
8192-bucket estimators, DLL4 is 12.9\(\times\) faster; at 8192
estimators, where neither data structure fits in cache, DLL4 is still
3.1\(\times\) faster.}
\end{figure}

\textbf{Cache thrashing (Figure 8).} When multiple estimator instances
are used simultaneously (as in per-\emph{k}-mer cardinality tracking),
LL6 suffers progressive cache thrashing as the combined working set
exceeds the CPU cache hierarchy. This single-threaded benchmark measures
pure \texttt{hashAndStore()} throughput with varying numbers of
simultaneous estimators on the AMD EPYC 7502P (32 KB L1d, 512 KB L2 per
core, 128 MB shared L3). With 512 simultaneous 8,192-bucket estimators
(total working set: 2 MB for DLL4, 4 MB for LL6 --- exceeding L2 but
fitting in L3), DLL4 achieves 2.3\(\times\) the throughput of LL6,
because eeMask eliminates \textgreater99.99\% of memory accesses
regardless of cache pressure. At the extreme (8,192 simultaneous
8,192-bucket estimators, total working set: 32--64 MB --- exceeding even
L3 in effective terms given the dual-socket CPU setup), DLL4 remains
3.1\(\times\) faster than LL6.

\hypertarget{low-complexity-evaluation}{%
\subsection{8.2 Low-Complexity
Evaluation}\label{low-complexity-evaluation}}

To test robustness under realistic conditions where many input elements
are duplicates, we used the low-complexity simulator with 256 buckets,
4,096 estimator instances, 8\(\times\) iterations, and cardinalities up
to 100 million (390,000\(\times\) the bucket count).

\begin{figure}
\centering
\includegraphics{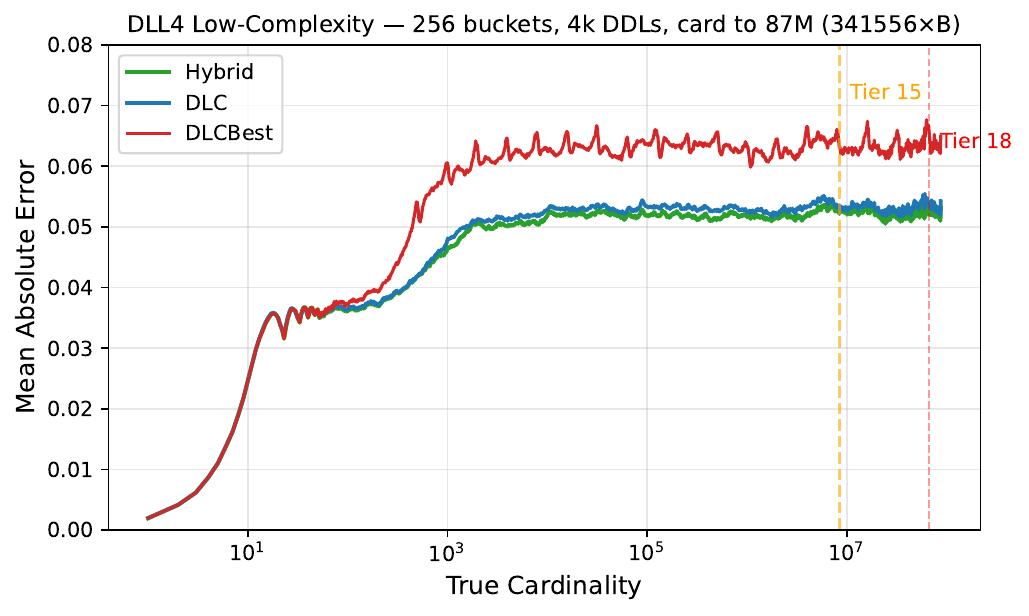}
\caption{\textbf{Figure 9.} DLL4 absolute error under low-complexity
conditions (256 buckets, 4k instances, 8\(\times\) iterations, up to
100M cardinality). Error remains flat through tier 15 and beyond --- the
4-bit range is sufficient for practical use.}
\end{figure}

\textbf{DLL4 stability (Figure 9).} DLL4 shows no error degradation at
any cardinality tested, including well past tier 16 (the point at which
all buckets may have, and most have, experienced overflow) and into tier
19. The error profile remains flat from low to extreme cardinality,
confirming that the 16-tier range of 4-bit storage is sufficient for
practical use. Buckets for these high-cardinality tests were restricted
to 256 to allow reaching high tiers with high precision, as the amount
of simulation time is
O(estimators\(\times\)buckets\(\times\)2\^{}maxTier). DLL4's Hybrid
achieves 4.83\% average and 5.30\% peak absolute error under
low-complexity conditions.

\begin{figure}
\centering
\includegraphics{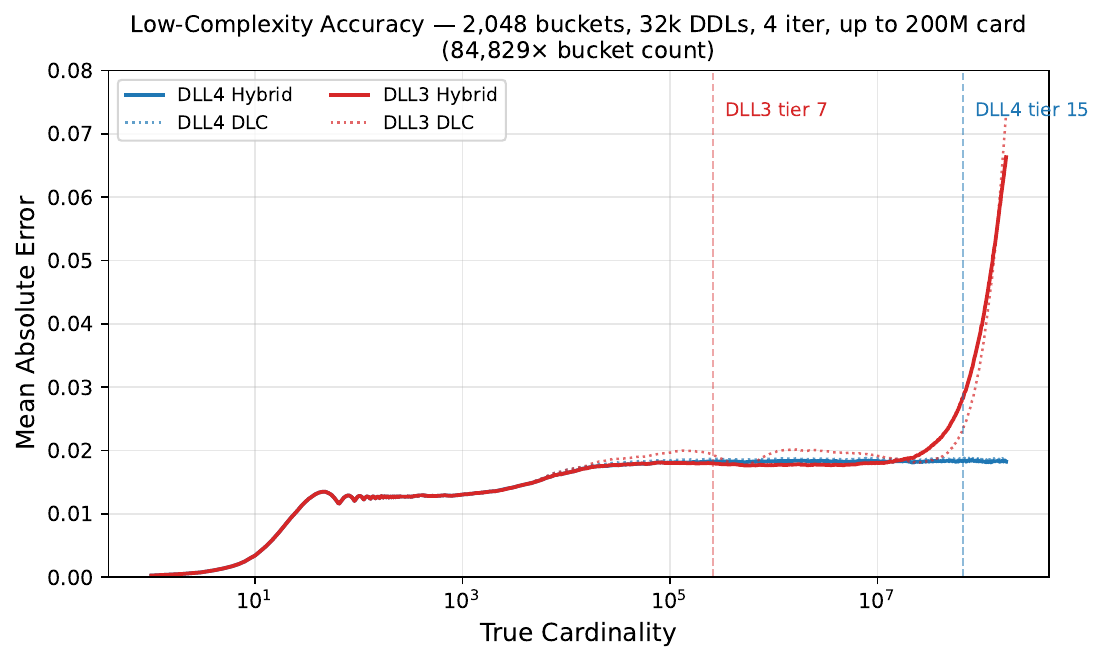}
\caption{\textbf{Figure 10.} DLL3 vs DLL4 absolute error under
low-complexity conditions at 2,048 buckets (32k instances, 4 iterations,
up to 200M cardinality = 84,829\(\times\) \emph{B}). DLL3 tracks DLL4
closely until tier 7 (\textasciitilde260k) then diverges sharply. DLL4
remains flat through tier 15 and beyond.}
\end{figure}

\begin{figure}
\centering
\includegraphics{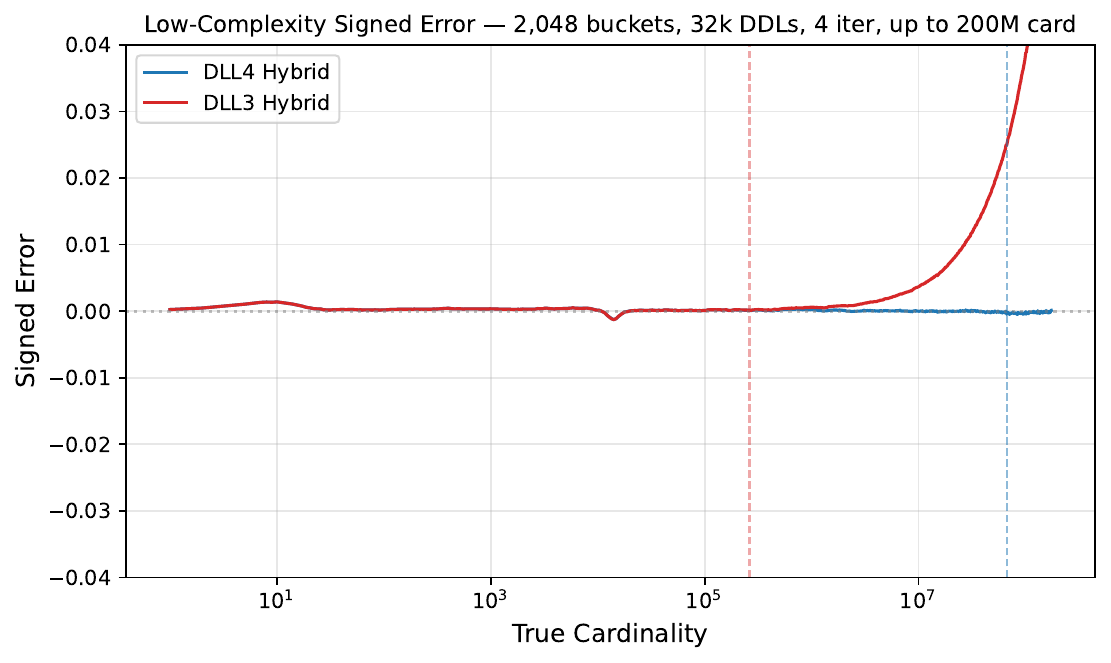}
\caption{\textbf{Figure 11.} DLL3 vs DLL4 signed error under
low-complexity conditions (same parameters as Figure 10). DLL3's
divergence above tier 7 is purely positive (overestimation); DLL4 stays
centered at zero throughout.}
\end{figure}

\textbf{DLL3 under low-complexity (Figures 10--11).} DLL3's behavior
under low-complexity data is nuanced and depends on both the
cardinality-to-bucket ratio and the duplication rate. At moderate
cardinalities, DLL3 shows comparable log-weighted average error to DLL4
--- Hybrid achieves 4.89\% (vs DLL4's 4.83\%) and DLC achieves 5.05\%
(vs 5.03\%). However, DLL3's peak error is moderately higher: Hybrid
peaks at 7.39\% (vs DLL4's 5.30\%).

The divergence onset shifts depending on simulation parameters. In our
100-million-cardinality test with 8\(\times\) iterations (reaching
\textasciitilde93 million true cardinality out of 100 million possible),
the error was still rising at the end of the simulation. With more
iterations (higher duplication rate) or closer approach to maximum
cardinality, the divergence would likely increase further. We report the
maximum observed divergence under our test conditions but cannot bound
the worst case, as the error depends on the specific duplication pattern
and how close the true cardinality approaches the array size.

Without overflow correction (Section 5.3), DLL3 would strictly
undercount in both high- and low-complexity cases, because clamping
always loses upward information. The overflow correction successfully
compensates for this in high-complexity data, where DLL3 matches DLL4
within 0.02\% under high-complexity conditions. However, under
low-complexity data with sustained duplication, the correction
overcorrects: overflowed duplicates that were already accounted for by
the correction can trigger additional tier promotions, causing the same
lost information to be compensated twice --- a systematic positive bias.
This overcorrection is responsible for DLL3's higher peak error (7.39\%
vs DLL4's 5.30\%) --- the divergence is caused \emph{by} the correction,
not despite it.

\textbf{Table 4.} Mean absolute error under low-complexity conditions
(256 buckets, 4k instances, 8\(\times\) iterations, up to 100M
cardinality). DLL3 matches DLL4 at moderate cardinality but diverges at
extreme duplication.

\begin{longtable}[]{@{}llll@{}}
\toprule
Estimator & Method & Avg Abs Error & Peak Abs Error\tabularnewline
\midrule
\endhead
DLL4 & Hybrid & 0.04833 & 0.05304\tabularnewline
DLL4 & DLC & 0.05028 & 0.05691\tabularnewline
DLL3 & Hybrid & 0.04892 & 0.07388\tabularnewline
DLL3 & DLC & 0.05051 & 0.06864\tabularnewline
\bottomrule
\end{longtable}

HLL (LL6) achieves 5.68\% average error under the same conditions ---
comparable to DLL4's DLC (5.03\%) --- but its peak error of 29.8\% is
5.6\(\times\) worse than DLL4's worst case (5.30\%), as the LC-to-HLL
transition spike is amplified under duplication.

\hypertarget{complementary-nature-of-dynamicloglog-and-ultraloglog}{%
\subsection{8.3 Complementary Nature of DynamicLogLog and
UltraLogLog}\label{complementary-nature-of-dynamicloglog-and-ultraloglog}}

UltraLogLog (ULL) {[}5{]} takes an orthogonal approach to improving
HLL's space efficiency: rather than sharing state to shrink registers
while maintaining information content, it stores additional sub-NLZ
history bits per register (2 extra bits recording whether updates with
NLZ$-1$ and NLZ$-2$ occurred), increasing both the size and
information content of each register. This enables more efficient
estimation via Ertl's FGRA (Further Generalized Remaining Area)
estimator, yielding a 28\% lower memory-variance product while
preserving HLL's independent-register architecture and full merge
semantics. Thus, both ULL and DLL increase the information density of
registers using opposite but compatible methods, as each method modifies
a different part of the register --- DLL increases the density of the
NLZ section, while ULL adds a new higher-density section. FGRA also
eliminates the LC-to-HLL transition spike: because it is a single
closed-form estimator that uses all register information simultaneously,
it has no discontinuous handoff between low-cardinality and
high-cardinality regimes (Figure 12). ULL does not, however, provide
early-exit acceleration --- every hash must be fully processed.

The two approaches are complementary. DLL's shared exponent requires
only 4 bits for relative NLZ, potentially freeing bits for sub-NLZ
history. We tested this directly by implementing UltraDynamicLogLog
(UDLL6): a 6-bit-per-register variant that combines DLL4's relative
encoding and tier promotion with ULL's 2-bit sub-NLZ history and FGRA
estimator. Each 6-bit register stores 4 bits of relative NLZ plus 2 bits
of history, in the same encoding as ULL but relative to the shared
exponent. This yields 2,048 registers in 1.5 KB --- or 1,640 bytes as
implemented, due to int-packing 5 registers per 32-bit word with 2 bits
unused per word (versus ULL's 2 KB for 2,048 8-bit registers, or DLL4's
1 KB for 2,048 4-bit registers).

\begin{figure}
\centering
\includegraphics{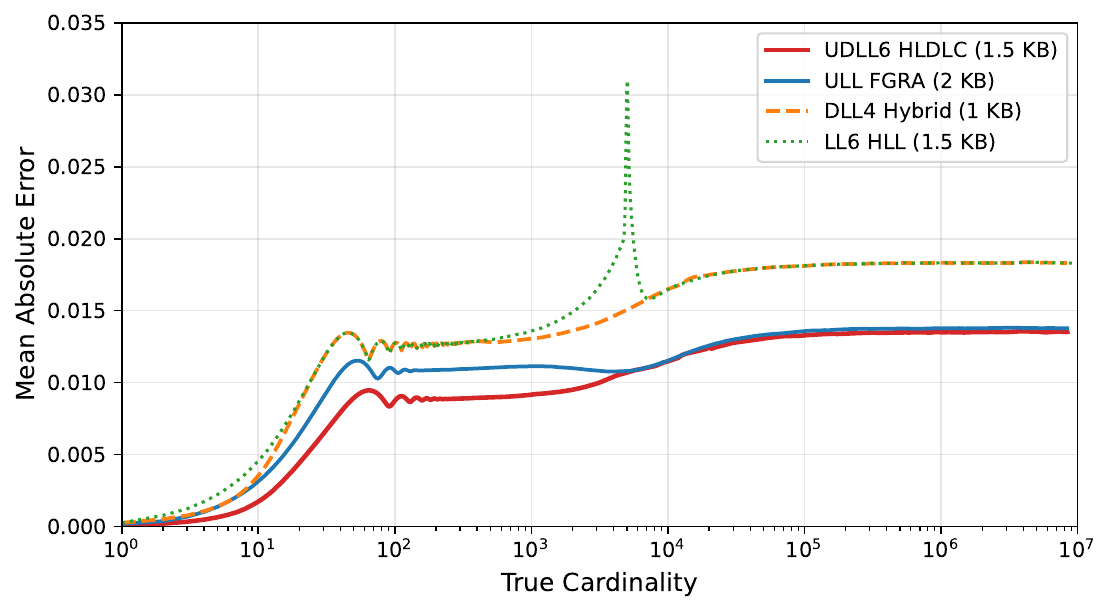}
\caption{\textbf{Figure 12.} Mean absolute error for DLL4, ULL (FGRA),
UDLL6 (HLDLC), and HLL (LL6) at 2,048 buckets (512k simulations, up to
8.4M cardinality). UDLL6 HLDLC at 1.5 KB achieves the lowest error
throughout --- 7\% better than ULL FGRA at 75\% of the memory. HLL's
transition bump peaks at 3.1\% near 5,000 cardinality (2.5\(\times\)
\emph{B}); DLL4 Hybrid and ULL FGRA eliminate it entirely.}
\end{figure}

Figure 12 and Table 3 compare accuracy across memory budgets. Two
averaging methods are shown: \emph{log-weighted} (unweighted average of
error at exponentially-spaced checkpoints, corresponding to what is
visible on a log-scale plot and reflecting an exponentially-distributed
cardinality workload) and \emph{width-weighted} (each checkpoint
weighted by its cardinality, emphasizing steady-state behavior at high
cardinality and corresponding to a linear-scale view).

For equal-memory comparisons at non-power-of-2 bucket counts, we use a
variant of DLL4 that selects buckets via unsigned modulo rather than
bitmask, allowing arbitrary bucket counts (e.g., 3,072 buckets for
exactly 1.5 KB at 4 bits per bucket). This uses the same estimator
pipeline and CF tables as the bitmask version.

\textbf{Table 3.} Mean absolute relative error at equal memory, averaged
to 8,388,608 cardinality. \emph{Log-wt} = unweighted average across
exponentially-spaced checkpoints; \emph{Width-wt} =
cardinality-interval-weighted average emphasizing steady state. HLL =
uncorrected LL6 (6-bit HyperLogLog) as baseline. DLC rows use no
correction factor; all other methods use CF where applicable.

\begin{longtable}[]{@{}lllllllll@{}}
\toprule
Estimator & Method & Memory & Buckets & Log-wt & vs HLL & Width-wt & vs
HLL & Peak\tabularnewline
\midrule
\endhead
LL6 & HLL & 1.5 KB & 2,048 & 0.01607 & --- & 0.01831 & --- &
0.03101\tabularnewline
DLL4 & Hybrid & 1 KB & 2,048 & 0.01572 & $-2\%$ & 0.01830 & 0\% &
0.01834\tabularnewline
DLL4 & DLC & 1 KB & 2,048 & 0.01617 & +1\% & 0.01898 & +4\% &
0.01928\tabularnewline
DLL4 & Hybrid & 1.5 KB & 3,072 & 0.01269 & $-21\%$ & 0.01494 &
$-18\%$ & 0.01498\tabularnewline
DLL4 & Hybrid & 2 KB & 4,096 & 0.01085 & $-32\%$ & 0.01294 & $-29\%$
& 0.01298\tabularnewline
DLL4 & DLC & 2 KB & 4,096 & 0.01114 & $-31\%$ & 0.01340 & $-27\%$ &
0.01373\tabularnewline
ULL & FGRA & 1 KB & 1,024 & 0.01741 & +8\% & 0.01949 & +6\% &
0.01955\tabularnewline
ULL & FGRA & 2 KB & 2,048 & 0.01210 & $-25\%$ & 0.01377 & $-25\%$ &
0.01381\tabularnewline
UDLL6 & FGRA & 1.5 KB & 2,048 & 0.01210 & $-25\%$ & 0.01376 &
$-25\%$ & 0.01380\tabularnewline
\bottomrule
\end{longtable}

At equal register count (2,048 buckets), DLL4 at 1 KB and HLL at 1.5 KB
have nearly identical error across all weightings (log-wt: $-2\%$,
width-wt: 0\%). The primary accuracy difference is in peak error: DLL4
Hybrid peaks at 1.83\% vs HLL's 3.10\%, eliminating the transition bump
entirely. When DLL4 is given the same memory as HLL (1.5 KB
\(\rightarrow\) 3,072 buckets via modulo addressing), the additional
buckets provide substantial improvement: $-21\%$ log-weighted,
$-18\%$ width-weighted.

UDLL6 at 1.5 KB and ULL at 2 KB overlap almost perfectly (Figure 12),
demonstrating that the fusion of DLL's shared exponent with ULL's
sub-NLZ history achieves ULL-level accuracy at 75\% of the memory. At
equal memory (1.5 KB), UDLL6 outperforms DLL4 by 4--8\%, suggesting that
FGRA's use of sub-NLZ history provides a small but consistent advantage
over the DLC/Hybrid pipeline at the same information budget.

ULL at 1 KB (1,024 registers) exhibits 6\% higher mean error than HLL at
1.5 KB in steady state despite its substantially more efficient
estimator --- FGRA extracts enough information from 1,024 8-bit
registers to nearly match 2,048 4-bit registers, but not quite. ULL's
improvement over HLL at equal register count is dramatic ($-25\%$
width-weighted at 2,048 buckets), confirming that FGRA is a major
advance. Yet DLL4 at equal \emph{memory} still outperforms ULL: the
additional buckets enabled by 4-bit storage provide a larger accuracy
gain than ULL's richer per-register information.

Adapting ULL's register encoding to DLL's relative framework required
rearranging the bit layout so that the NLZ portion occupies the high
bits of each register, enabling the same unsigned-comparison early exit
used by DLL4. However, because UDLL6 must accept hashes whose NLZ falls
within the 2-bit history margin below the current floor (to update
sub-NLZ history for registers near the floor), its eeMask threshold is
necessarily 2 bits more conservative than DLL4's. This reduces the early
exit rate, which may partially explain UDLL6's lower throughput.

In production, UDLL6 is 54\% faster than ULL (1,388 vs 901 Mbp/s in
BBDuk; Figure 7) and nearly matches DLL4 with eeMask (1,415 Mbp/s). The
speed gain over ULL comes from int-packed 6-bit registers (5 per 32-bit
word, 1,640 bytes for 2,048 registers) combined with the conservative
eeMask that rejects most hashes before any register access.

\hypertarget{history-corrected-hybrid-estimation}{%
\subsection{8.4 History-Corrected Hybrid
Estimation}\label{history-corrected-hybrid-estimation}}

UltraLogLog's per-register history bits (Section 8.3) record whether
updates with NLZ$-1$ and NLZ$-2$ occurred, providing sub-NLZ
information that FGRA exploits via Fisher-information analysis. An
alternative to FGRA is direct per-state correction: each of the
2\^{}\emph{h} history states (where \emph{h} is the number of history
bits) exhibits a characteristic bias relative to the tier average, and
this bias can be measured by simulation and stored as a small additive
correction table. We call the resulting estimators
\textbf{Hybrid+\emph{n}} (or \textbf{Mean+\emph{n}}), where \emph{n} is
the number of history bits used.

For 2-bit history, each register falls into one of 4 states based on
whether NLZ$-1$ and NLZ$-2$ updates have been observed. Simulation
over 6.5 million trials per state yields a 4-entry additive correction
in NLZ-space (e.g., state 0 contributes $-2.51$ to the effective NLZ,
while state 3 contributes +0.21). These per-state corrections are
applied before the harmonic mean computation, making each register a
less biased estimator of its local cardinality. The approach generalizes
naturally to 1-bit and 3-bit history (2 and 8 correction entries,
respectively). We denote the resulting estimator types by their total
bits per register: \textbf{UDLL5} (4 NLZ + 1 history), \textbf{UDLL6} (4
NLZ + 2 history), and \textbf{UDLL7} (4 NLZ + 3 history). Note that
UDLL6 is architecturally identical to the UDLL6 described in Section 8.3
--- the same data structure supports both FGRA and Hybrid+2 as
estimation methods.

\begin{figure}
\centering
\includegraphics{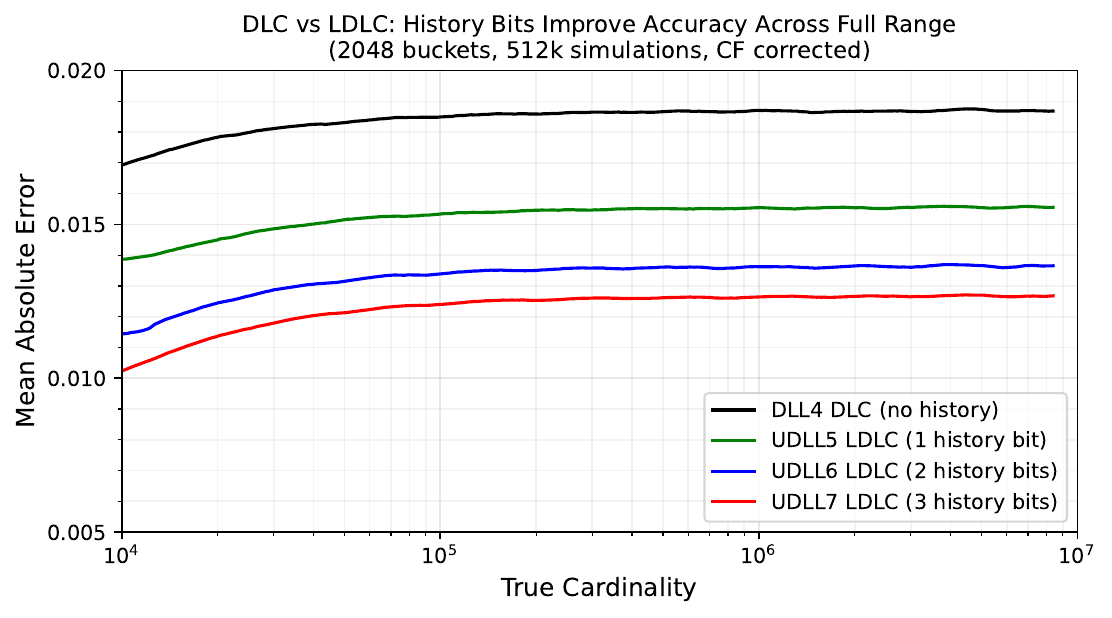}
\caption{\textbf{Figure 13.} LDLC accuracy by history bits: DLL4 DLC
(0-bit baseline), UDLL5 LDLC (1-bit), UDLL6 LDLC (2-bit), and UDLL7 LDLC
(3-bit), all at 2,048 buckets with CF correction. History bits improve
LDLC accuracy across the full cardinality range: the first bit reduces
log-weighted error by 12\%, the second by an additional 19\%, and the
third by a further 15\% (comparing within each implementation). (512k
simulations.)}
\end{figure}

\textbf{Table 5.} Mean absolute error by estimation method (2,048
buckets, CF-corrected, 512k simulations). \emph{Log-wt} = unweighted
average across exponentially-spaced checkpoints; \emph{Width-wt} =
cardinality-interval-weighted average.

\begin{longtable}[]{@{}lllll@{}}
\toprule
Estimator & Method & Bits/reg & Log-wt & Width-wt\tabularnewline
\midrule
\endhead
DLL4 & DLC & 4 & 0.01588 & 0.01868\tabularnewline
UDLL5 & LDLC & 5 & 0.01400 & 0.01554\tabularnewline
UDLL6 & LDLC & 6 & 0.01128 & 0.01363\tabularnewline
UDLL7 & LDLC & 7 & 0.01192 & 0.01265\tabularnewline
UDLL6 & FGRA & 6 & 0.01210 & 0.01376\tabularnewline
\bottomrule
\end{longtable}

Each additional history bit improves LDLC accuracy across the full
cardinality range (Figure 13): the first bit reduces log-weighted error
by 12\%, the second by 19\%, and the third by 15\%. At 2 history bits,
UDLL6 LDLC (log-weighted 0.01128) outperforms UDLL6 FGRA (0.01210) using
the DLC pipeline with history-only linear counting, compared to FGRA's
full Fisher-information analysis with \(\sigma\)/\(\varphi\) series
coefficients.

At equal memory, history bits outperform extra buckets:

\textbf{Table 6.} Memory-fair comparison: UDLL LDLC vs DLL4 Hybrid at
equal memory (512k simulations).

\begin{longtable}[]{@{}lllllll@{}}
\toprule
Memory (bytes) & UDLL LDLC & Log-wt & Width-wt & DLL4 Hybrid & Log-wt &
Width-wt\tabularnewline
\midrule
\endhead
1,368 & UDLL5 (2,048\emph{B}) & \textbf{0.01400} & \textbf{0.01554} &
DLL4 (2,560\emph{B}) & 0.01394 & 0.01637\tabularnewline
1,640 & UDLL6 (2,048\emph{B}) & \textbf{0.01128} & \textbf{0.01363} &
DLL4 (3,072\emph{B}) & 0.01266 & 0.01494\tabularnewline
1,824 & UDLL7 (2,048\emph{B}) & 0.01192 & \textbf{0.01265} & DLL4
(3,584\emph{B}) & \textbf{0.01163} & 0.01382\tabularnewline
\bottomrule
\end{longtable}

UDLL LDLC wins on both weighting schemes at 2 history bits: UDLL6 LDLC
is 11\% better log-weighted and 9\% better width-weighted than DLL4
Hybrid at equal memory. At 1 and 3 history bits, LDLC wins on
width-weighted error (where history bits improve steady-state accuracy)
while DLL4 is competitive on log-weighted error. UDLL6 offers the best
practical tradeoff: 93.7\% packing efficiency in 32-bit words and the
lowest error of any configuration tested.

\hypertarget{layered-dynamic-linear-counting-ldlc}{%
\subsection{8.5 Layered Dynamic Linear Counting
(LDLC)}\label{layered-dynamic-linear-counting-ldlc}}

The per-state history corrections can also improve DLC. At each tier,
the cumulative DLC empty count \emph{V\_t} can be refined by
incorporating History-only Linear Counting (HC): an LC estimate computed
using only the history bits of registers at the current tier boundary.
DLC and HC exhibit systematic biases that are complementary --- where
DLC overestimates at a given cardinality, HC tends to underestimate, and
vice versa (Figure 14, upper panel).

\begin{figure}
\centering
\includegraphics{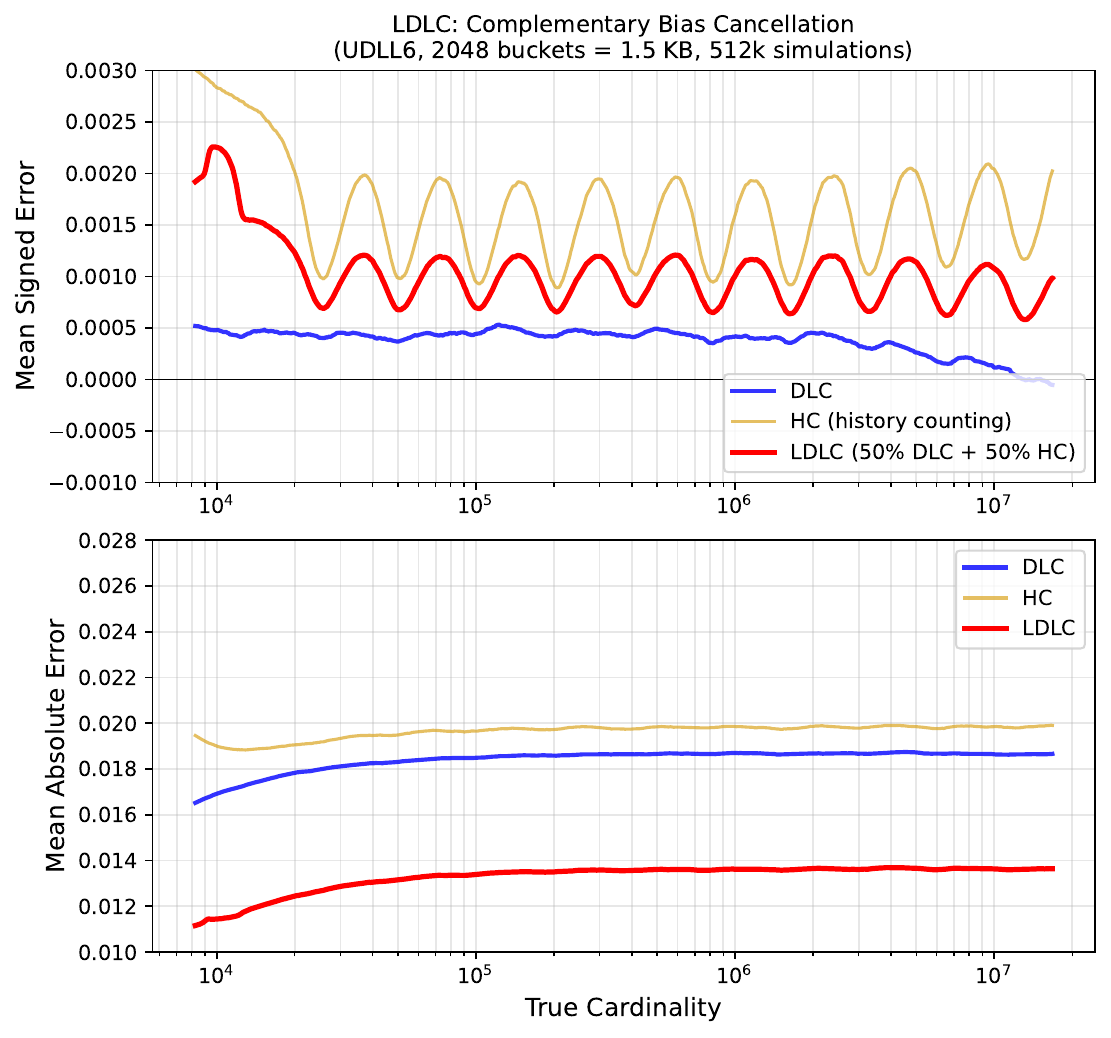}
\caption{\textbf{Figure 14.} Complementary bias cancellation in LDLC.
Upper panel: signed error of DLC (blue) and HC (gold) shows opposite
periodic structure; the LDLC blend (red) threads between them. Lower
panel: absolute error showing LDLC's 27\% width-weighted improvement
over DLC. (UDLL6, 2,048 buckets, 1.5 KB, 512k simulations.)}
\end{figure}

LDLC blends these two estimates as a weighted sum: 50\% DLC + 50\% HC.
The complementary cancellation reduces the periodic bias component,
yielding a 27\% improvement in width-weighted error over DLC alone
(1.36\% vs 1.87\% at 2,048 buckets). Like DLC, LDLC requires no
correction factor table --- the accuracy gain comes entirely from
combining two complementary estimators that are individually unbiased on
average but biased in opposite directions at any given cardinality.

\begin{figure}
\centering
\includegraphics{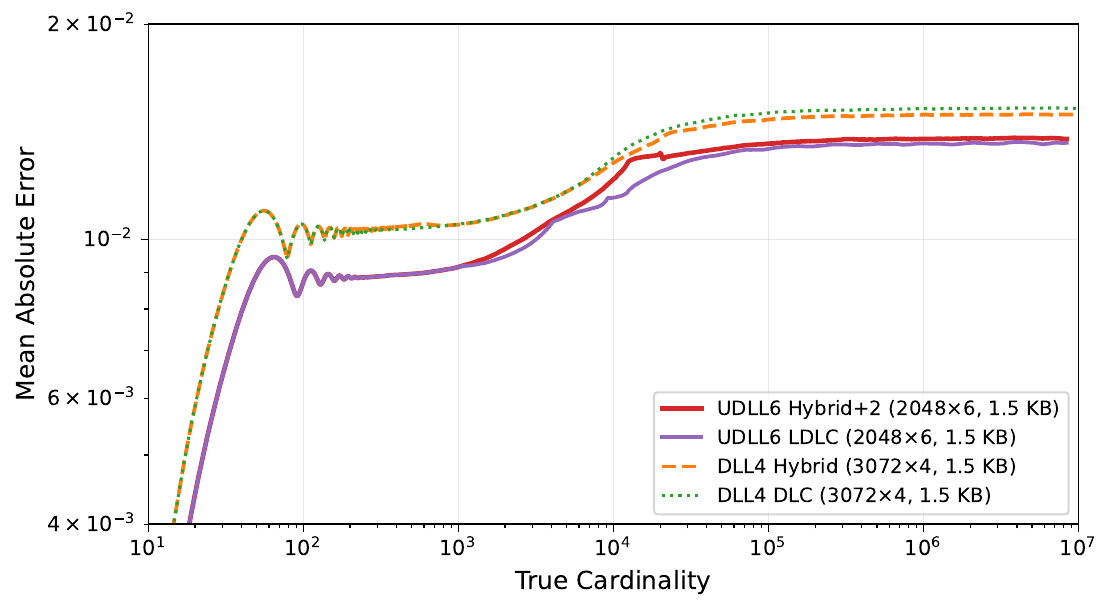}
\caption{\textbf{Figure 15.} Memory-fair comparison at 1.5 KB: DLL4
Hybrid and DLC (3,072 \(\times\) 4-bit), UDLL6 Hybrid+2, LDLC, and FGRA
(2,048 \(\times\) 6-bit). All estimators eliminate the LC-to-HLL
transition artifact; history-corrected methods achieve the lowest error.
(512k simulations.)}
\end{figure}

\textbf{HLDLC: Hybrid LDLC.} LDLC and Hybrid+2 are complementary: LDLC
derives its accuracy from DLC's tier-aware LC framework with HC bias
cancellation, while Hybrid+2 derives its accuracy from CF-corrected Mean
with per-state history corrections. Because these two estimators use
different mathematical pipelines --- one LC-based, the other
harmonic-mean-based --- their residual errors are largely uncorrelated.
HLDLC exploits this by blending them equally: HLDLC = 0.5 \(\times\)
LDLC + 0.5 \(\times\) Hybrid+2. This simple average achieves
\textbf{1.12\%} log-weighted and \textbf{1.35\%} width-weighted mean
absolute error at 2,048 buckets (1.5 KB) --- a 7\% improvement over FGRA
(1.21\% / 1.38\%) at 75\% of the memory. HLDLC is currently the most
accurate estimator tested in our framework.

\hypertarget{comparison-with-external-estimators}{%
\subsection{8.6 Comparison with External
Estimators}\label{comparison-with-external-estimators}}

To place DLL and UDLL6 in context, we benchmarked them against
reimplementations of two widely used external cardinality estimators
under identical conditions (512,000 instances, 2,048 buckets,
cardinality up to 8,388,608).

\textbf{Apache DataSketches HLL\_4} {[}6{]} uses 4-bit registers
relative to a global \texttt{curMin} offset, with values exceeding the
4-bit range stored in an auxiliary hash map. The architecture is
structurally similar to DLL4's shared exponent, but the exception map
adds variable memory overhead beyond the base 1 KB register array. We
reimplemented HLL\_4 from the algorithm description (clean-room, no
library dependency).

\textbf{HyperLogLogLog (HLLL)} {[}7{]} uses 3-bit registers relative to
a global base with a sparse exception map for overflows, yielding a base
footprint of 768 bytes for 2,048 registers. Periodic rebasing minimizes
exception count. We ported the reference C++ implementation to Java,
using BBTools' \texttt{IntHashMap} for the exception map.

Both external estimators use the standard HyperLogLog harmonic mean with
small-range linear counting correction --- they do not have access to
DLL's DLC, Hybrid, or history-corrected estimators. Each estimator's
best available estimation method was used.

\begin{figure}
\centering
\includegraphics{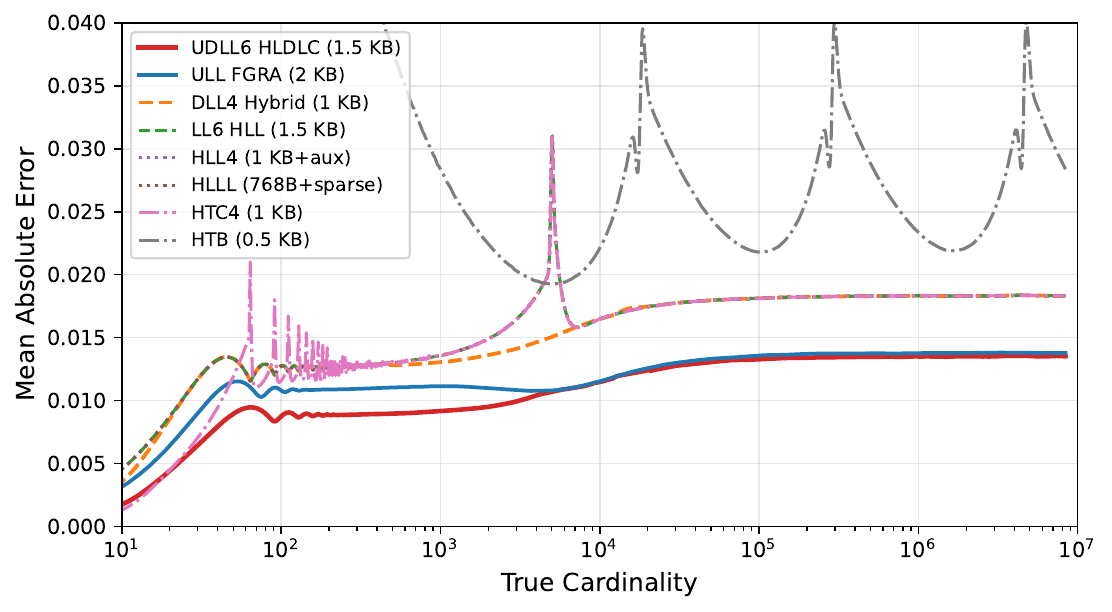}
\caption{\textbf{Figure 16.} Mean absolute error for all eight estimator
types from Table 7 at 2,048 buckets (512k simulations, up to 8.4M
cardinality). UDLL6 HLDLC at 1.5 KB achieves the lowest error
throughout. HLL\_4, HLLL, and HTC4 exhibit the characteristic HLL
transition spike near 2--5\(\times\) \emph{B}; DLL variants eliminate it
entirely. HTB (0.5 KB) shows high error at low cardinality.}
\end{figure}

\begin{longtable}[]{@{}llllllll@{}}
\caption{\textbf{Table 7.} Comparison of estimator types at 2,048
buckets (512k simulations, cardinality up to 8,388,608). \emph{Log-wt} =
unweighted average across exponentially-spaced checkpoints;
\emph{Width-wt} = cardinality-interval-weighted average. Speed columns:
per-core million adds/second on AMD EPYC 7502P (128 threads on a
128-core machine, divided by 128); \emph{1/thr} = single estimator per
thread (ALU-limited); \emph{4096/thr} = 4,096 simultaneous estimators
per thread (bandwidth-limited).}\tabularnewline
\toprule
Type & Method & Memory & Log-wt & Width-wt & Peak & Speed (1/thr) &
Speed (4096/thr)\tabularnewline
\midrule
\endfirsthead
\toprule
Type & Method & Memory & Log-wt & Width-wt & Peak & Speed (1/thr) &
Speed (4096/thr)\tabularnewline
\midrule
\endhead
UDLL6 & HLDLC & 1.5 KB & \textbf{0.01123} & \textbf{0.01351} & 0.01356 &
90.0 & 38.1\tabularnewline
ULL (Ertl) & FGRA & 2 KB & 0.01211 & 0.01377 & 0.01381 & 60.1 &
11.9\tabularnewline
DLL4 & Hybrid & 1 KB & 0.01567 & 0.01832 & 0.01837 & 96.3 &
111.2\tabularnewline
LL6 (HLL) & HLL & 1.5 KB & 0.01607 & 0.01831 & 0.03101 & 78.0 &
12.7\tabularnewline
HLL\_4 (Apache) & HLL & 1 KB+aux & 0.01606 & 0.01831 & 0.03101 & 53.6 &
13.3\tabularnewline
HLLL (Karppa) & HLL & 768B+sparse & 0.01606 & 0.01831 & 0.03101 & 76.3 &
40.5\tabularnewline
HTC4 (Xiao) & HTC & 1 KB & 0.01594 & 0.01831 & 0.03096 & 80.3 &
44.3\tabularnewline
HTB (Sedgewick) & HTB & 0.5 KB & 0.04059 & 0.02861 & 1.00000 & 149.8 &
123.7\tabularnewline
\bottomrule
\end{longtable}

DLL4 at 1 KB matches LL6 at 1.5 KB in steady-state accuracy (1.83\%
width-weighted), confirming that the shared-exponent architecture
recovers the accuracy lost by reducing from 6-bit to 4-bit registers.
HLL\_4 and HLLL --- despite different register widths and exception-map
architectures --- produce identical accuracy (1.61\% log-weighted,
1.83\% width-weighted), because both use the same HLL harmonic mean
estimator with LC transition. Their 3.1\% peak error (vs DLL4's 1.84\%)
is the HLL transition spike that DLC eliminates.

HTC4 (HLL-TailCut+) {[}9{]} uses a shared-offset architecture similar to
DLL4. Its steady-state accuracy (1.83\% width-weighted) is comparable to
HLL\_4 and HLLL, with a similar 3.1\% peak at the LC-to-HLL transition.
HTC4 is not idempotent. HTB (HyperTwoBits) {[}11{]} uses only 2 bits per
register (0.5 KB at 2,048 buckets) and achieves 2.86\% width-weighted
error in the steady-state region, but exhibits 100\% peak error at very
low cardinality (below \textasciitilde10\(\times\) \emph{B}).

The speed columns reveal a striking architectural divide. In the
ALU-limited regime (1 estimator/thread), all types achieve 54--150 M
adds/s per core, with HTB fastest (149.8) due to its minimal 0.5 KB
footprint and DLL4 second (96.3). Under bandwidth pressure (4,096
simultaneous estimators/thread), types without an early-exit mechanism
--- ULL, LL6, HLL\_4 --- collapse 5--6\(\times\) as every add hits DRAM.
DLL4, by contrast, \emph{accelerates} to 111.2 M/s because its
early-exit mask (eeMask) skips memory access entirely when the incoming
NLZ falls below the global floor; at high cardinality, over 99\% of adds
early-exit. HTB exhibits the same behavior at half the memory (123.7 M/s
at 4k/thr). HTC4 and HLLL show intermediate resilience (44.3 and 40.5
M/s respectively), likely due to their shared-offset and sparse-encoding
architectures reducing effective working-set size.

Among all estimators tested, UDLL6 HLDLC achieves the lowest
width-weighted error (1.35\%) at 1.5 KB --- 25\% less memory than ULL's
2 KB, and 26\% lower error than standard HLL estimation used by HLL\_4,
HLLL, and HTC4.

\hypertarget{discussion}{%
\section{9. Discussion}\label{discussion}}

DynamicLogLog demonstrates that the LogLog framework's memory and speed
can be substantially improved by sharing state across buckets. The
shared-exponent design is conceptually simple --- analogous to
floating-point representation with a global absolute exponent and
per-bucket relative exponent offsets --- but its consequences are
far-reaching: 33\% memory savings, up to 12.9\(\times\) higher
throughput in bandwidth-limited workloads (Figure 8), and the
elimination of the LC-to-HLL transition artifact.

The DLL family of estimators is perhaps the most practically significant
contribution. Where HLL requires a carefully tuned transition between
two fundamentally different estimators (LC and harmonic mean), DLC
provides a single, unified framework that is accurate across the full
cardinality range. The tier-aware extension of LC is natural given DLL's
architecture, but it could in principle be applied to any LogLog variant
that tracks per-bucket NLZ values.

The per-state history correction mechanism (Mean+H) uses 16 state-table
entries (4 steady-state corrections plus 3 tiers \(\times\) 4 states) in
addition to a 16-coefficient cardinality-keyed correction formula (3S2G)
--- 32 constants total --- to capture most of the information that FGRA
extracts through full Fisher-information analysis. LDLC further shows
that complementary bias cancellation between DLC and history-only LC can
improve accuracy without any correction tables at all.

A mantissa variant, DynamicDemiLog (DDL), trades bucket count for
per-bucket collision resistance by storing fractional NLZ bits. This
enables set comparison operations analogous to MinHash sketching, as
well as element frequency histograms. These capabilities will be
described in a forthcoming paper.

\textbf{Idempotency.} DLL is not idempotent: adding the same element
twice can produce a different state than adding it once, because tier
promotion caused by intervening elements may bring a previously-capped
value back into the representable range (Section 5.4). This property is
shared by other shared-offset approaches including the tailcut method
{[}9{]} and the HyperBitBit family {[}11{]}. In contrast, HyperLogLog
and UltraLogLog are fully idempotent --- each element's contribution
depends only on its hash, not on insertion order or multiplicity. The
practical impact of non-idempotency depends on the workload. For streams
of unique elements (the high-complexity case), idempotency is mostly
irrelevant since no duplicates occur, though different orderings can
cause slightly different estimates. For streams with duplicates (the
low-complexity case), Section 8.2 demonstrates that DLL4 maintains
stable accuracy (4.83\% average, 5.30\% peak) under sustained
duplication, and the error profile remains flat across the full
cardinality range. The primary consequence of DLL's non-idempotent
architecture is in merge operations, discussed below.

\textbf{Limitations.}

\begin{figure}
\centering
\includegraphics{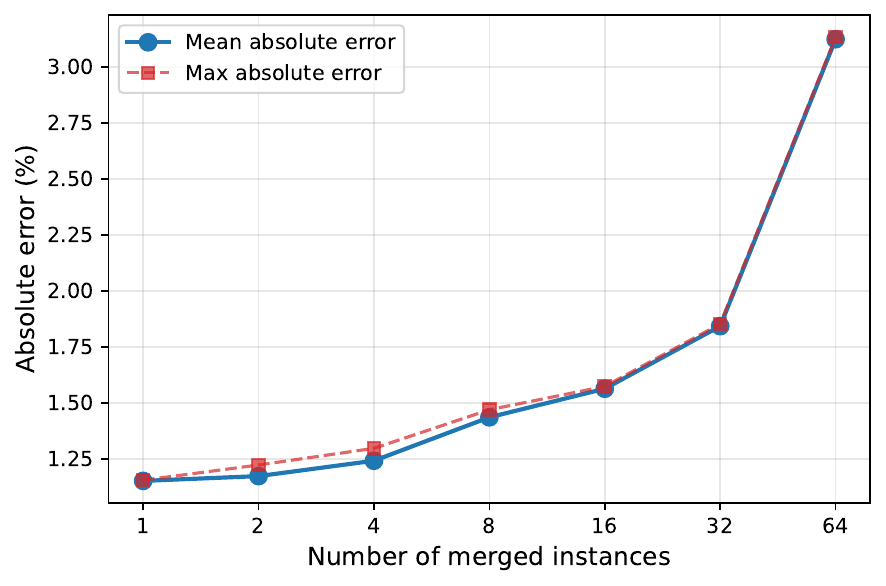}
\caption{\textbf{Figure 17.} Merge accuracy: mean absolute error vs
number of merged DLL4 instances, at true cardinality 1.72 billion with
2048 buckets. Error is negligible at 8 instances (0.02\%) and grows to
1.4\% at 64 instances.}
\end{figure}

DLL's shared-exponent design introduces a tradeoff for parallel merge
operations. When multiple DLL instances process disjoint subsets of a
stream and are merged (via per-bucket max), each instance promotes tiers
less aggressively than a single instance seeing all data. This increases
the effective overflow at merge boundaries, producing a systematic
underestimate. With DLL4's 16-tier range, the effect is minor in
practice --- approximately 0.02\% undercount with 8 merged instances,
0.15\% with 16, and 1.4\% with 64, on a 1.72-billion-element dataset
with 2,048 buckets. For parallel cardinality estimation of a single
stream, we recommend using a single synchronized DLL instance rather
than per-thread copies. Traditional HLL, which has no tier promotion,
does not suffer from this issue.

\textbf{Future work.} The eeMask early exit technique is applicable to
any LogLog-family estimator that adopts tier promotion, and could be
retroactively added to existing implementations. Additionally, the
self-similar CF structure (Section 5.2) generalizes to any estimator
with periodic error structure. The merge undercount could potentially be
addressed by a merge compensation factor derived from the number of
merged instances and the tier gap between them, though this remains to
be investigated. ExaLogLog {[}10{]}, which achieves the best known
memory-variance product among fixed-register HLL variants, uses
variable-width registers (6+\emph{t}+\emph{d} bits) that are
substantially larger per register than DLL4's 4 bits --- at 2,048
buckets, each ExaLogLog register occupies approximately 32 bits (8 KB
total vs DLL4's 1 KB). A direct comparison at equal memory would require
either reducing ExaLogLog's register count or increasing DLL4's; this
comparison is deferred to a forthcoming paper focused on the 1 KB memory
regime.

\hypertarget{conclusion}{%
\section{10. Conclusion}\label{conclusion}}

DynamicLogLog achieves higher accuracy and up to 12.9\(\times\) higher
throughput per element than HyperLogLog at 33\% less memory. The key
innovations --- shared-exponent storage, early exit masking, and Dynamic
Linear Counting --- are individually simple but collectively
transformative. DLL provides a drop-in replacement for HyperLogLog in
any application where cardinality estimation is needed, with significant
advantages in accuracy, speed, and memory, and only a minor disadvantage
in merge accuracy for highly parallel workloads (Section 9).
Furthermore, DLL's framework is complementary with UltraLogLog's
per-register history approach: their fusion (UDLL6) achieves ULL-level
accuracy at 75\% of the memory, and per-state history corrections
(Hybrid+2) nearly match FGRA using only a 10-entry correction table.
Layered Dynamic Linear Counting (LDLC) extends this further, achieving
27\% better width-weighted accuracy than DLC through complementary bias
cancellation --- with no correction tables at all. HLDLC, a simple
average of LDLC and Hybrid+2, achieves the lowest error of any estimator
tested (1.12\% log-weighted, 1.35\% width-weighted at 2,048 buckets, 1.5
KB) --- 7\% better than ULL's FGRA at 25\% less memory, and 26\% better
than standard HLL estimation as used by Apache DataSketches HLL\_4 and
HyperLogLogLog. These results demonstrate that architectural compaction
and information-theoretic efficiency are orthogonal improvements that
compose naturally.

\hypertarget{acknowledgements}{%
\section{Acknowledgements}\label{acknowledgements}}

Simulations and benchmarks were conducted on the Dori computing cluster
at the Joint Genome Institute, Lawrence Berkeley National Laboratory.
The work conducted by the U.S. Department of Energy Joint Genome
Institute (https://ror.org/04xm1d337), a DOE Office of Science User
Facility, is supported by the Office of Science of the U.S. Department
of Energy operated under Contract No.~DE-AC02-05CH11231. Synthetic
benchmark reads were generated using \texttt{randomgenome.sh} and
\texttt{randomreadsmg.sh} from the BBTools suite and are fully
reproducible by any user with those tools.

The author thanks Alex Copeland for reviewing and commenting on the
manuscript, Otmar Ertl and Jim Apple for providing feedback on the
preprint, and Chloe for substantial contributions to algorithm
implementation, calibration infrastructure, and manuscript preparation
throughout this project.

\textbf{Disclosure of AI assistance.} This work made extensive use of
large language model (LLM) AI tools, including Anthropic Claude and
Google Gemini. The calibration drivers, benchmarking infrastructure, and
supporting analysis scripts described in the Methods were developed with
AI assistance. The initial draft of this manuscript, including all
sections, was prepared with AI assistance and subsequently reviewed,
revised, and validated by the author. All algorithmic design decisions,
experimental methodology, interpretation of results, and scientific
conclusions are solely those of the author, who takes full
responsibility for the accuracy of all content. AI tools were used as an
instrument of productivity, not as a source of scientific judgment. This
disclosure is provided in accordance with emerging journal policies on
AI-assisted authorship and in the interest of full transparency.

\hypertarget{references}{%
\section*{References}\label{references}}
\addcontentsline{toc}{section}{References}

\hypertarget{refs}{}
\begin{cslreferences}
\leavevmode\hypertarget{ref-whang1990}{}%
{[}1{]} K.-Y. Whang, B. T. Vander-Zanden, and H. M. Taylor, ``A
linear-time probabilistic counting algorithm for database
applications,'' \emph{ACM Transactions on Database Systems}, vol. 15,
no. 2, pp. 208--229, 1990, doi:
\href{https://doi.org/10.1145/78922.78925}{10.1145/78922.78925}.

\leavevmode\hypertarget{ref-durand2003}{}%
{[}2{]} M. Durand and P. Flajolet, ``Loglog counting of large
cardinalities,'' in \emph{Algorithms -- esa 2003}, in Lecture notes in
computer science, vol. 2832. Springer, 2003, pp. 605--617. doi:
\href{https://doi.org/10.1007/978-3-540-39658-1_55}{10.1007/978-3-540-39658-1\_55}.

\leavevmode\hypertarget{ref-flajolet2007}{}%
{[}3{]} P. Flajolet, \'E. Fusy, O. Gandouet, and F. Meunier,
``HyperLogLog: The analysis of a near-optimal cardinality estimation
algorithm,'' in \emph{Proceedings of the 2007 international conference
on analysis of algorithms (aofa)}, in Discrete mathematics and
theoretical computer science. 2007, pp. 137--156.

\leavevmode\hypertarget{ref-heule2013}{}%
{[}4{]} S. Heule, M. Nunkesser, and A. Hall, ``HyperLogLog in practice:
Algorithmic engineering of a state of the art cardinality estimation
algorithm,'' \emph{Proceedings of the EDBT 2013 Conference}, pp.
683--692, 2013, doi:
\href{https://doi.org/10.1145/2452376.2452456}{10.1145/2452376.2452456}.

\leavevmode\hypertarget{ref-ertl2024}{}%
{[}5{]} O. Ertl, ``UltraLogLog: A practical and more space-efficient
alternative to HyperLogLog for approximate distinct counting,''
\emph{Proceedings of the VLDB Endowment}, vol. 17, no. 7, pp.
1655--1668, 2024, doi:
\href{https://doi.org/10.14778/3654621.3654632}{10.14778/3654621.3654632}.

\leavevmode\hypertarget{ref-datasketches}{}%
{[}6{]} Apache Software Foundation, ``Apache DataSketches: A software
library of stochastic streaming algorithms.'' 2024. Available:
\url{https://github.com/apache/datasketches-java}

\leavevmode\hypertarget{ref-karppa2022}{}%
{[}7{]} M. Karppa and R. Pagh, ``HyperLogLogLog: Cardinality estimation
with one log more,'' in \emph{Proceedings of the 28th acm sigkdd
conference on knowledge discovery and data mining}, 2022, pp. 753--761.
doi:
\href{https://doi.org/10.1145/3534678.3539246}{10.1145/3534678.3539246}.

\leavevmode\hypertarget{ref-flajolet1985}{}%
{[}8{]} P. Flajolet and G. N. Martin, ``Probabilistic counting
algorithms for data base applications,'' \emph{Journal of Computer and
System Sciences}, vol. 31, no. 2, pp. 182--209, 1985, doi:
\href{https://doi.org/10.1016/0022-0000(85)90041-8}{10.1016/0022-0000(85)90041-8}.

\leavevmode\hypertarget{ref-xiao2017}{}%
{[}9{]} Q. Xiao, Y. Zhou, and S. Chen, ``Better with fewer bits:
Improving the performance of cardinality estimation of large data
streams,'' in \emph{IEEE infocom 2017}, 2017, pp. 1--9. doi:
\href{https://doi.org/10.1109/INFOCOM.2017.8057088}{10.1109/INFOCOM.2017.8057088}.

\leavevmode\hypertarget{ref-ertl2025exa}{}%
{[}10{]} O. Ertl, ``ExaLogLog: Space-efficient and practical approximate
distinct counting up to the exa-scale,'' in \emph{Proceedings of the
28th international conference on extending database technology (edbt)},
2025. Available:
\url{https://openproceedings.org/2025/conf/edbt/paper-252.pdf}

\leavevmode\hypertarget{ref-sedgewick2013}{}%
{[}11{]} S. Janson, J. Lumbroso, and R. Sedgewick, ``Bit-array-based
alternatives to HyperLogLog,'' in \emph{Proceedings of the 35th
international conference on probabilistic, combinatorial and asymptotic
methods for the analysis of algorithms (aofa 2024)}, in Leibniz
international proceedings in informatics (lipics), vol. 302. Schloss
Dagstuhl -- Leibniz-Zentrum für Informatik, 2024, pp. 5:1--5:19. doi:
\href{https://doi.org/10.4230/LIPIcs.AofA.2024.5}{10.4230/LIPIcs.AofA.2024.5}.

\leavevmode\hypertarget{ref-ertl2017}{}%
{[}12{]} O. Ertl, ``New cardinality estimation algorithms for
HyperLogLog sketches,'' in \emph{ArXiv preprint}, 2017. Available:
\url{https://arxiv.org/abs/1702.01284}

\leavevmode\hypertarget{ref-bbtools}{}%
{[}13{]} B. Bushnell, ``BBTools: Bioinformatics tools for sequence
analysis.'' 2014. Available: \url{https://bbmap.org}

\leavevmode\hypertarget{ref-wang1997}{}%
{[}14{]} T. Wang, ``Integer hash function.'' 1997. Available:
\url{https://web.archive.org/web/2007/http://www.concentric.net/~Ttwang/tech/inthash.htm}

\leavevmode\hypertarget{ref-bloom1970}{}%
{[}15{]} B. H. Bloom, ``Space/time trade-offs in hash coding with
allowable errors,'' \emph{Communications of the ACM}, vol. 13, no. 7,
pp. 422--426, 1970, doi:
\href{https://doi.org/10.1145/362686.362692}{10.1145/362686.362692}.

\leavevmode\hypertarget{ref-blackman2021}{}%
{[}16{]} D. Blackman and S. Vigna, ``Scrambled linear pseudorandom
number generators,'' \emph{ACM Transactions on Mathematical Software},
vol. 47, no. 4, pp. 1--32, 2021, doi:
\href{https://doi.org/10.1145/3460772}{10.1145/3460772}.
\end{cslreferences}

\end{document}